\documentclass{aa}
\usepackage{natbib}
\bibpunct{(}{)}{;}{a}{}{,}

\usepackage[dvips]{graphicx}
\usepackage{times}
\usepackage{rotating}
\usepackage{longtable}
\newcommand\ignore[1]{} 

\begin{document}
\title{
Reprocessing the Hipparcos Intermediate Astrometric Data of spectroscopic binaries: II. Systems with a giant component
\thanks{Based on observations from the Hipparcos
astrometric satellite operated by the European Space Agency (ESA 1997) and on data collected with the Simbad database}}
\titlerunning{Re-processing the Hipparcos IAD of spectroscopic binaries. II.} 
\author{D.~Pourbaix\inst{1}\fnmsep\thanks{Research Associate, F.N.R.S., Belgium}
\and 
H.M.J.~Boffin\inst{2}}
\institute{
Institut d'Astronomie et d'Astrophysique, Universit\'e Libre de Bruxelles, C.P.~226, Boulevard du Triomphe, B-1050 Bruxelles, Belgium 
\and 
Observatoire Royal de Belgique, Avenue circulaire 3, B-1180 Bruxelles, Belgium}
\date{Received date; accepted date} 
\offprints{pourbaix@astro.ulb.ac.be}
\abstract{
By reanalyzing the Hipparcos Intermediate Astrometric Data of a large sample of spectroscopic binaries containing 
a giant, we obtain a sample of 29 systems
fulfilling a carefully derived set of constraints and hence for which we can derive an accurate orbital solution.
Of these, one is a double-lined spectroscopic binary and six were not listed in the DMSA/O section of the catalogue. 
Using our solutions, we derive the masses of the components in these systems and statistically analyze them.
We also briefly discuss each system individually.
}
\maketitle

%
\section{Introduction}
%
Barium stars and other extrinsic peculiar red giants (PRGs) are now, almost without doubt, believed
to result from their binarity \citep{Boffin-1988:a,Jorissen-1992:a}:
their companion, presumably a white dwarf \citep[e.g.][]{BohmVitense-2000:a}, transfered
through its stellar wind and when on the Asymptotic Giant Branch (AGB), matter enriched in carbon and s-process elements. 
The level of contamination of the barium star will therefore depend on several factors, 
some depending on the orbital properties of the system as well as on the stellar wind velocities \citep{Theuns-1996:a,Mastrodemos-1998:a,Nagae-2002:a}. 
Because a large quantity of matter has been lost by the system when the companion evolved from
the AGB phase to its present white dwarf nature, the present orbital properties of barium stars 
are not the original ones. In particular, for most systems (with the exception of the smallest), 
the orbital separation must be larger while the eccentricity should remain more or less constant
\citep{Theuns-1996:a,Jorissen-1992:a,Boffin-1994:a,Karakas-2000:a,Liu-2000:a}.

There also exists a class of spectroscopic red giant binaries whose orbits have similar characteristics
to those of the barium stars yet they show normal abundances \citep{Jorissen-1992:a,Boffin-1993:a,Zacs-1997:a}. 
Hence, the binarity is a necessary condition for barium stars but not a sufficient one.

\citet[][~BCP93 in the following]{Boffin-1993:a} performed a statistical analysis of a sample of 
spectroscopic binaries containing late-type (spectral type G or K) giants but not showing 
the chemical anomalies of barium stars. 
Their sample of 213 systems served as a comparison for the sample of known peculiar red giant stars. They 
derived the distribution of the orbital elements, studied the eccentricity-period relation and used two
different methods to obtain the mass ratio distribution, which they concluded was in good agreement 
with a uniform mass ratio distribution function and a single-valued giant mass of 1.5 M$_\odot$ 
or with a distribution peaked towards small mass ratios for a giant mass of
3 M$_\odot$.

In this paper, we follow-up on the analysis of \citet{Boffin-1993:a} by taking advantage of the
results of the ESA astrometric space mission Hipparcos \citep{Hipparcos}. We have thus extended the sample
of BCP93 by trying to add all orbits published between 1993 and the end of 2001 of systems containing a G-K giant and not showing the chemical anomalies
of PRGs. The resulting sample for which Hipparcos
data were available contains in total 215 stars, listed in Table~\ref{tab:allsample}. 
For all these stars, we have reanalyzed the Hipparcos Intermediate Astrometric Data (IAD) and
assessed the reliability of the astrometric binary solution following \citet{Pourbaix-2001:b}. We then applied 
a statistical test of the periodogram and obtain 29 systems for which we are confident
that the astrometric solution based on the IAD is fully consistent with the
spectroscopic orbit. This is certainly a very conservative approach but it has the merit
of not trying to fit some "noise" in the data.
For these 29 systems, we can make use of the knowledge of the parallax ($\varpi$) and their effective 
temperature to estimate their location in an Hertzsprung-Russel diagram and by comparison with evolutionary
tracks to estimate the mass of the giant primary. With the knowledge of the inclination of the system,
we can then estimate the mass ratio and hence the mass of the unseen companion. Apart from gaining 
useful insight into a few selected spectroscopic systems, this methodology should allow to 
obtain a more accurate estimate of the masses and mass ratio distribution of spectroscopic binaries
containing red giants and compare with the results of BCP93 and other studies.

Section \ref{Sect:orbits} presents our analysis of the astrometric orbits and the selection of the 29 systems. The masses of the primary are then inferred from astrophysical and statistical considerations (Sect.~\ref{Sect:massfuncprim}).   In Sect.~\ref{Sect:masssec}, the masses of the secondary are derived from the primary masses and the inclinations and their distribution is investigated.  The only double-lined system we keep is analyzed in Sect.~\ref{Sect:sb2} whereas the results for each individual system are briefly summarized in Sect.~\ref{Sect:individual}.

%
\section{Astrometric orbits}\label{Sect:orbits}
%

The Hipparcos observations of all the known spectroscopic binaries were not
fitted with an orbital model.  Some of them were processed as single stars
(5 parameters) while others were assigned a stochastic solution.  
\citet{Pourbaix-2000:b} fitted the IAD and
derived an astrometric orbit for about eighty spectroscopic binaries 
processed as single stars in the Hipparcos Catalogue.  Some physical 
assumptions (e.g. mass of the primary) were then used to assess the 
reliability of these orbits and only 25\% of them were finally accepted.

Instead of these physical assumptions, one can also fit the data with two
distinct sets of orbital parameters and assess the reliability of the
orbit from the agreement between the two solutions \citep{Pourbaix-2001:b}.
Let us briefly summarize the two approaches keeping in mind that in both cases, the eccentricity, orbital period and periastron time are assumed from the spectroscopic orbit.  On the one hand, the four remaining parameters, i.e. the semi-major axis of the photocentric orbit ($a_0$), the inclination ($i$), the latitude of the ascending node ($\Omega$) and the argument of the periastron ($\omega$) are fitted through their Thiele-Innes constant combination.  On the other hand, two more parameters are assumed from the spectroscopic orbit, namely the amplitude of the radial velocity curve and $\omega$.  Here, only two parameters of the photocentric orbit ($i$ and $\Omega$) are thus derived.  Although that double-fit approach was designed in the context of the astrometric orbits of extra solar planets, it can be applied to spectroscopic binaries, especially single-lined.

%
\subsection{First screening of the astrometric solutions}\label{Sect:accepted}
%
Among the 215 spectroscopic binaries investigated (Table \ref{tab:allsample}), the fit of the IAD is significantly improved with an orbital model \citep[][F-test at 5\%,]{Pourbaix-2001:a} for 152 of them.  However, only 100 have fitted orbital parameters (Thiele-Innes' constants) which are significantly non zero.

\begin{table*}[hbt]
\caption[]{\label{tab:allsample}List of all the 215 stars investigated in this study}
\setlength{\tabcolsep}{1.9mm}
\begin{tabular}{rr rr rr rr rr rr r}\hline
HIP & HIP & HIP & HIP & HIP & HIP & HIP & HIP & HIP & HIP & HIP & HIP & HIP \\ \hline
\object{\ignore{HIP }443}&\object{\ignore{HIP }664}&\object{\ignore{HIP }759}&\object{\ignore{HIP }1792}&\object{\ignore{HIP }2081}&\object{\ignore{HIP }2170}&\object{\ignore{HIP }2900}&\object{\ignore{HIP }3092}&\object{\ignore{HIP }3494}&\object{\ignore{HIP }4463}&\object{\ignore{HIP }5744}&\object{\ignore{HIP }5951}&\object{\ignore{HIP }7143}\\
\object{\ignore{HIP }7487}&\object{\ignore{HIP }7719}&\object{\ignore{HIP }8086}&\object{\ignore{HIP }8645}&\object{\ignore{HIP }8833}&\object{\ignore{HIP }8922}&\object{\ignore{HIP }9631}&\object{\ignore{HIP }10280}&\object{\ignore{HIP }10324}&\object{\ignore{HIP }10340}&\object{\ignore{HIP }10366}&\object{\ignore{HIP }10514}&\object{\ignore{HIP }10969}\\
\object{\ignore{HIP }11304}&\object{\ignore{HIP }11784}&\object{\ignore{HIP }12488}&\object{\ignore{HIP }13531}&\object{\ignore{HIP }14328}&\object{\ignore{HIP }14763}&\object{\ignore{HIP }15041}&\object{\ignore{HIP }15264}&\object{\ignore{HIP }15807}&\object{\ignore{HIP }15900}&\object{\ignore{HIP }16369}&\object{\ignore{HIP }17136}&\object{\ignore{HIP }17440}\\
\object{\ignore{HIP }17587}&\object{\ignore{HIP }17932}&\object{\ignore{HIP }18782}&\object{\ignore{HIP }20455}&\object{\ignore{HIP }21144}&\object{\ignore{HIP }21476}&\object{\ignore{HIP }21727}&\object{\ignore{HIP }22176}&\object{\ignore{HIP }23743}&\object{\ignore{HIP }24085}&\object{\ignore{HIP }24286}&\object{\ignore{HIP }24331}&\object{\ignore{HIP }24608}\\
\object{\ignore{HIP }24727}&\object{\ignore{HIP }26001}&\object{\ignore{HIP }26714}&\object{\ignore{HIP }26795}&\object{\ignore{HIP }26953}&\object{\ignore{HIP }27588}&\object{\ignore{HIP }28343}&\object{\ignore{HIP }28734}&\object{\ignore{HIP }29071}&\object{\ignore{HIP }29982}&\object{\ignore{HIP }30501}&\object{\ignore{HIP }30595}&\object{\ignore{HIP }31019}\\
\object{\ignore{HIP }31062}&\object{\ignore{HIP }32578}&\object{\ignore{HIP }32768}&\object{\ignore{HIP }35600}&\object{\ignore{HIP }35658}&\object{\ignore{HIP }36284}&\object{\ignore{HIP }36377}&\object{\ignore{HIP }36690}&\object{\ignore{HIP }36992}&\object{\ignore{HIP }37629}&\object{\ignore{HIP }37908}&\object{\ignore{HIP }39424}&\object{\ignore{HIP }40326}\\
\object{\ignore{HIP }40470}&\object{\ignore{HIP }40772}&\object{\ignore{HIP }41939}&\object{\ignore{HIP }42432}&\object{\ignore{HIP }42673}&\object{\ignore{HIP }43109}&\object{\ignore{HIP }43903}&\object{\ignore{HIP }44946}&\object{\ignore{HIP }45527}&\object{\ignore{HIP }45875}&\object{\ignore{HIP }46168}&\object{\ignore{HIP }46893}&\object{\ignore{HIP }47205}\\
\object{\ignore{HIP }47206}&\object{\ignore{HIP }49841}&\object{\ignore{HIP }50109}&\object{\ignore{HIP }52032}&\object{\ignore{HIP }52085}&\object{\ignore{HIP }53240}&\object{\ignore{HIP }54632}&\object{\ignore{HIP }56862}&\object{\ignore{HIP }57565}&\object{\ignore{HIP }57791}&\object{\ignore{HIP }59148}&\object{\ignore{HIP }59459}&\object{\ignore{HIP }59600}\\
\object{\ignore{HIP }59736}&\object{\ignore{HIP }59796}&\object{\ignore{HIP }59856}&\object{\ignore{HIP }60364}&\object{\ignore{HIP }60555}&\object{\ignore{HIP }61724}&\object{\ignore{HIP }62886}&\object{\ignore{HIP }62915}&\object{\ignore{HIP }63613}&\object{\ignore{HIP }65187}&\object{\ignore{HIP }65417}&\object{\ignore{HIP }66286}&\object{\ignore{HIP }66358}\\
\object{\ignore{HIP }66907}&\object{\ignore{HIP }67013}&\object{\ignore{HIP }67234}&\object{\ignore{HIP }67480}&\object{\ignore{HIP }67615}&\object{\ignore{HIP }67744}&\object{\ignore{HIP }68180}&\object{\ignore{HIP }69112}&\object{\ignore{HIP }69879}&\object{\ignore{HIP }71332}&\object{\ignore{HIP }73721}&\object{\ignore{HIP }75233}&\object{\ignore{HIP }75325}\\
\object{\ignore{HIP }75356}&\object{\ignore{HIP }75989}&\object{\ignore{HIP }76196}&\object{\ignore{HIP }78322}&\object{\ignore{HIP }78985}&\object{\ignore{HIP }79195}&\object{\ignore{HIP }79345}&\object{\ignore{HIP }79358}&\object{\ignore{HIP }80166}&\object{\ignore{HIP }80816}&\object{\ignore{HIP }82080}&\object{\ignore{HIP }83336}&\object{\ignore{HIP }83575}\\
\object{\ignore{HIP }83947}&\object{\ignore{HIP }84014}&\object{\ignore{HIP }84291}&\object{\ignore{HIP }84677}&\object{\ignore{HIP }85680}&\object{\ignore{HIP }85749}&\object{\ignore{HIP }86579}&\object{\ignore{HIP }86946}&\object{\ignore{HIP }87428}&\object{\ignore{HIP }87472}&\object{\ignore{HIP }88696}&\object{\ignore{HIP }89860}&\object{\ignore{HIP }90135}\\
\object{\ignore{HIP }90313}&\object{\ignore{HIP }90441}&\object{\ignore{HIP }90659}&\object{\ignore{HIP }90692}&\object{\ignore{HIP }91636}&\object{\ignore{HIP }91751}&\object{\ignore{HIP }91784}&\object{\ignore{HIP }91820}&\object{\ignore{HIP }92512}&\object{\ignore{HIP }92550}&\object{\ignore{HIP }92782}&\object{\ignore{HIP }92818}&\object{\ignore{HIP }92872}\\
\object{\ignore{HIP }93244}&\object{\ignore{HIP }93305}&\object{\ignore{HIP }95066}&\object{\ignore{HIP }95244}&\object{\ignore{HIP }95342}&\object{\ignore{HIP }95714}&\object{\ignore{HIP }95808}&\object{\ignore{HIP }96003}&\object{\ignore{HIP }96467}&\object{\ignore{HIP }96683}&\object{\ignore{HIP }96714}&\object{\ignore{HIP }97384}&\object{\ignore{HIP }97979}\\
\object{\ignore{HIP }98351}&\object{\ignore{HIP }99011}&\object{\ignore{HIP }99847}&\object{\ignore{HIP }100361}&\object{\ignore{HIP }100764}&\object{\ignore{HIP }101098}&\object{\ignore{HIP }101847}&\object{\ignore{HIP }101953}&\object{\ignore{HIP }102388}&\object{\ignore{HIP }103356}&\object{\ignore{HIP }103519}&\object{\ignore{HIP }103890}&\object{\ignore{HIP }104684}\\
\object{\ignore{HIP }104894}&\object{\ignore{HIP }104987}&\object{\ignore{HIP }105017}&\object{\ignore{HIP }105583}&\object{\ignore{HIP }106013}&\object{\ignore{HIP }106241}&\object{\ignore{HIP }106497}&\object{\ignore{HIP }107089}&\object{\ignore{HIP }109002}&\object{\ignore{HIP }109303}&\object{\ignore{HIP }110130}&\object{\ignore{HIP }111072}&\object{\ignore{HIP }111171}\\
\object{\ignore{HIP }112158}&\object{\ignore{HIP }112997}&\object{\ignore{HIP }113478}&\object{\ignore{HIP }114222}&\object{\ignore{HIP }114421}&\object{\ignore{HIP }116278}&\object{\ignore{HIP }116584}&&&&&& \\
\hline
\end{tabular}
\end{table*}

The assessment scheme after \citet{Pourbaix-2001:b} is essentially based on
the improvement of the fit with the orbital model with respect to the single
star one ($Pr_2\le 5\%$), the significance of the Thiele-Innes constants 
resulting from the fit ($Pr_3\le 5\%$), the consistency of the Thiele-Innes 
solution and the Campbell/spectroscopic one ($Pr_4\ge 5\%$) and the likelihood 
of the face-on orbit ($Pr_5\ge 5\%$).  There are still 51 systems that 
simultaneously fulfill these four criteria, i.e. systems whose astrometric 
orbit can presumably be trusted.

\begin{figure}[htb]
\resizebox{\hsize}{!}{\includegraphics[angle=-90]{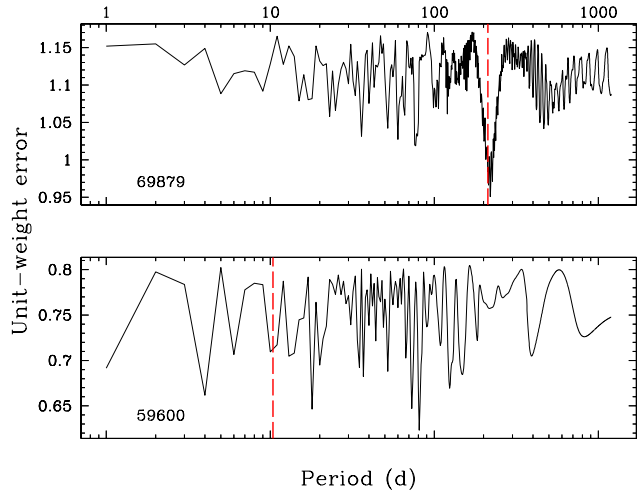}}
\caption[]{\label{fig:periodogram}Periodogram and the orbital period (dashed line) for HIP 69879 and 59600.  In the former, there is a peak at the expected period, thus showing that what Hipparcos saw is the system we investigate.  In the latter, there is no peak around 10 days.  A mistake in the orbital period as well as an additional companion are still possible.}
\end{figure}

\begin{figure*}[htb]
\resizebox{0.5\hsize}{!}{\includegraphics{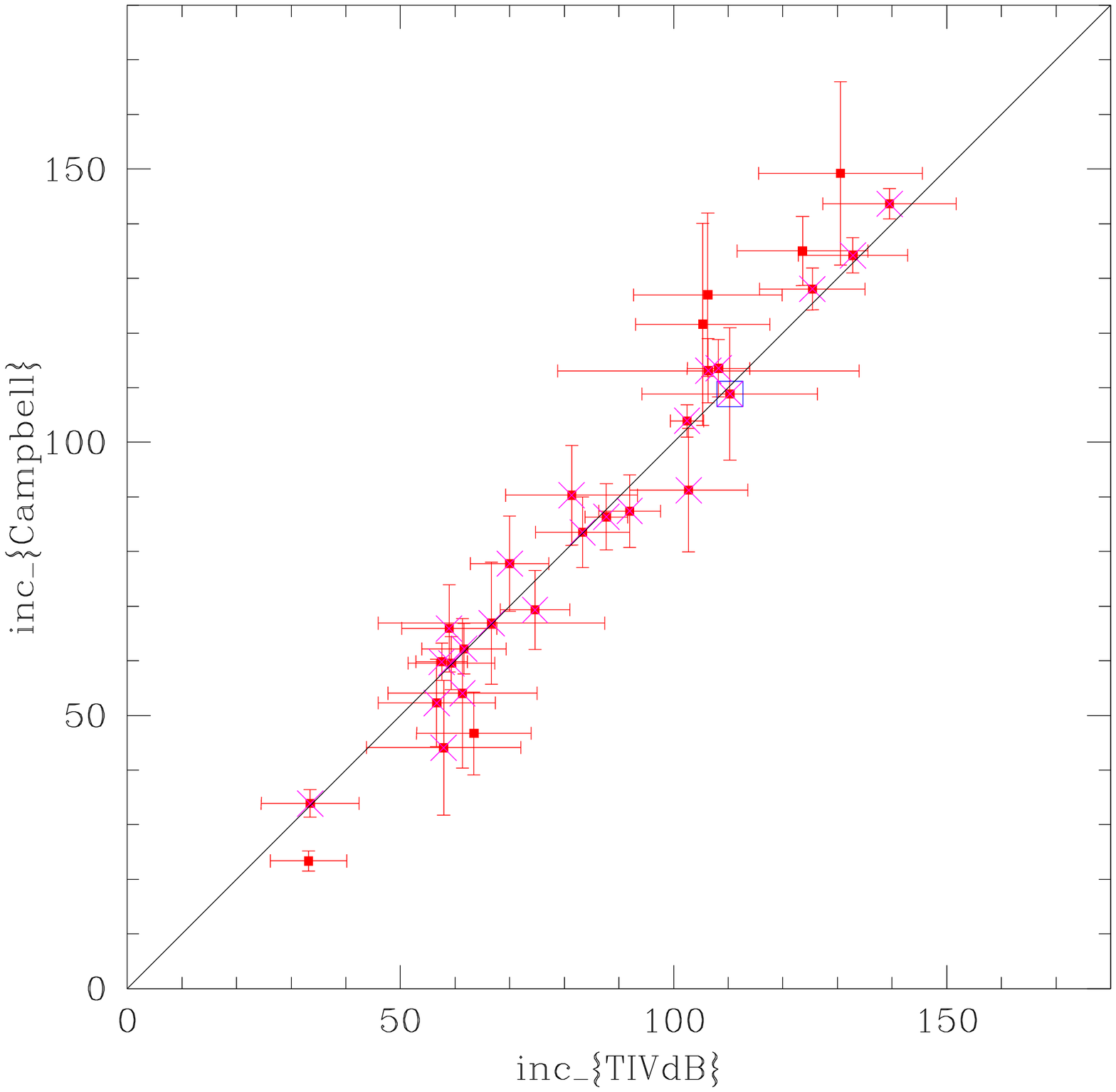}}\hfill
\resizebox{0.5\hsize}{!}{\includegraphics{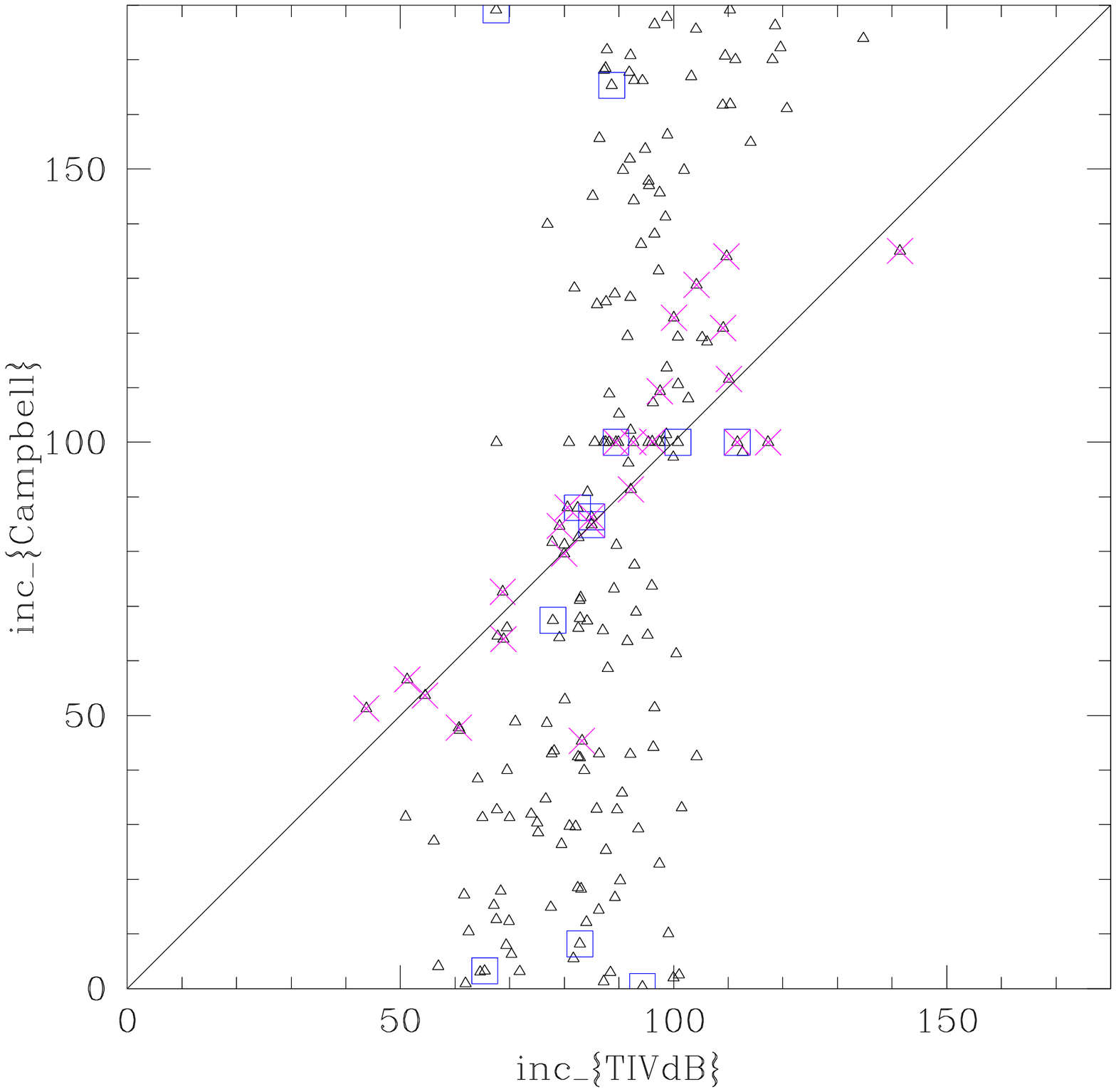}}
\caption[]{\label{fig:inclinations}Comparison of the inclinations derived from the Thiele-Innes constant and directly as one of the Campbell elements.  The left panel represents the 29 accepted systems whereas the right one gives the 190 rejected ones.  Crosses denote DMSA/O objects.  Open boxes denote double-lined spectroscopic binaries.}
\end{figure*}

Actually, these tests are based on the fit only and fulfilling them does not mean that what Hipparcos saw is the companion we are interested in and for which we partially adopted an orbital solution.  A periodogram can thus be used to see if the IAD do contain a peak at the period corresponding to the spectroscopic one.  Two examples are given in Fig.~\ref{fig:periodogram}.  On the one hand, the periodogram of HIP~69879 exhibits a very deep peak at the right location where, on the other hand, the IAD of HIP~59600 do not show any special behavior around 10 days.

Besides these criteria on the periodicity in the signal and the statistical assessment of the fit, one can also add a criterion based on the correlation among the model parameters (e.g., the efficiency $\epsilon$), say $\epsilon\ge0.4$ \citep{Eichhorn-1989} and another about a S/N threshold below which no reliable solution is expected, say $a_0\sin i\ge 1.5$ mas.  Combining all these constraints, one ends up with 29 systems only.  The resulting inclinations are plotted in the left panel of Fig.~\ref{fig:inclinations}.  One notices that the systems fulfilling those constraints also have inclinations derived with the two methods in fairly good agreement.  

On the other hand, the distribution of the inclinations for the rejected systems is, at first sight, more puzzling.  Indeed, the $i$ fitted directly with Campbell's approach are almost uniformly distributed over $[0, 180\degr]$ whereas those resulting from the inversion of the Thiele-Innes constants remain clustered around 90$\degr$.  Such clustering is typical of systems where the Thiele-Innes constants are non significantly different from zero as shown by \citet{Pourbaix-2001:b} in the context of extra-solar planets.  The inclinations derived from Campbell's method were there always close to 0 or $180\degr$ (whereas they are here uniform over $[0, 180\degr]$).  However, face-on orbits are not per se the sign of absence of any orbital astrometric signal.  \citet{Pourbaix-2001:a} showed that, at low S/N,
\[
\sin i\sim\frac{a_0\sin i}{\sigma_{\rm IAD}} <a_0\sin i
\]
the right-hand side of the expression coming from the spectroscopic orbit.  In this relation, $a_0\sin i$ is the projected angular size of the spectroscopic orbit (the secondary is assumed to be unseen) and $\sigma_{\rm IAD}$ is the standard deviation of the abscissa residuals.  Whereas, in case of extra-solar planets, $a_0\sin i$ is tiny, it gets very large with spectroscopic binaries (assuming the same units).  Therefore, $\sin i$ is no longer constrained to be nearly 0 and $i$ and, thus, can get (uniformly) distributed over $[0, 180\degr]$.  The right panel of Fig.~\ref{fig:inclinations} is thus a perfect illustration of the inclinations derived with the two approaches when there is no astrometric wobble.

Another point worth noting is that a lot of the systems previously processed with an orbital solution (DMSA/O) are no longer kept as such.  
This is also true for most of the SB2 systems, as all but one had to be discarded.  
However, these SB2 systems are often rejected because of the small difference of magnitude between the components thus leading to a tiny semi-major axis of the photocentric orbit.  On the other hand, about 30\% of the systems for which we adopt an orbital solution were initially processed as single stars.
It should also be noted that for the majority of the systems we keep and for which an orbital solution (DMSA/O) exists, the orbital 
elements we derive are generally of much greater accuracy. This makes it useful for the analysis which follows.

\begin{figure}[htb]
\resizebox{\hsize}{!}{\includegraphics{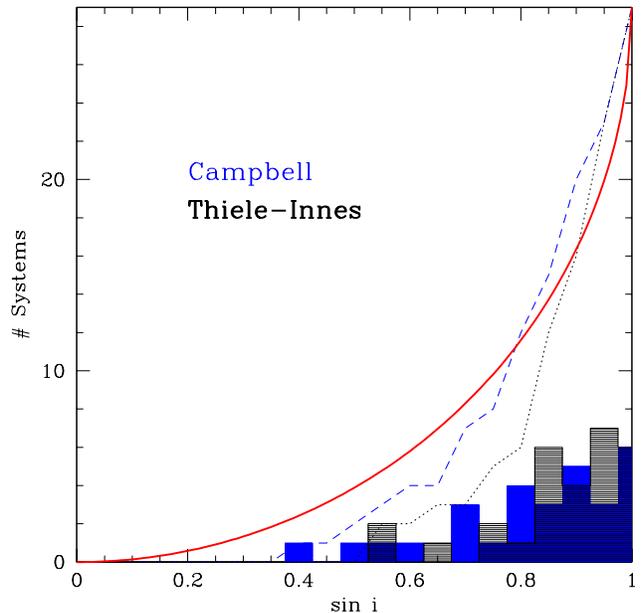}}
\caption[]{\label{fig:distinc}Distribution of the inclinations derived with the two methods.  The grey (black) histogram gives the number of inclinations in bins of 0.05 obtained with Campbell's (Thiele-Innes') method.  The dashed (dotted) line are the corresponding cumulative distributions whereas the solid line represents the distribution if the orbits are randomly oriented.  The small excess of nearly edge-on orbits results from a selection effect.
}
\end{figure}

Table \ref{tab:inclinations} gives the inclinations of the 29 accepted orbits and their distribution is given in Fig~\ref{fig:distinc}.  This distribution is quite consistent with randomly oriented orbits, with just a small excess of nearly edge-on orbits which is an expected observational bias for spectroscopic binaries.  Even if the Hipparcos observations are quite precise and accurate, the average precision on these inclination is about 8\degr.  With a behavior in $\sin i$, the consequence of such an uncertainty on the derived mass becomes more important the closer one gets to face-on orbits.
For example, even if for the system HIP~8922, the one-sigma error on the inclination is below two degrees, the resulting
error on the mass function is larger than 20\%. And sometimes, the error on the mass function can become larger than 50\%.
For a few systems, however, mostly these with an inclination close to 90$\degr$, the error on the mass function can be 
of the order of 2\%. As for these systems, we have generally also a very accurate parallax, the physical parameters 
we can derive will prove very significant.
It is noteworthy that five systems have inclinations close to 90$\degr$, hence which could show eclipses (Table~\ref{tab:eclipses}). As the companion
is supposedly a main sequence star and the separations are rather large, the conditions for eclipses to occur are 
stringent and the eventual eclipses  will be very partial, i.e. very weak. 
It may in fact be more appropraite to talk of transits.
The systems being however very bright, 
even the small change in luminosity such eclipse would produce should easily be observable by a careful amateur astronomer.
Because of the error on the orbital period, it is however not possible to
predict with any reliability the time of conjonction.

\begin{table}[htb]
\caption[]{\label{tab:inclinations}Inclination of the orbits of the 29 kept systems}
\setlength{\tabcolsep}{1.9mm}
\begin{tabular}{lc|lc|lc}\hline
HIP & $i$ (\degr) & HIP & $i$ (\degr) & HIP & $i$ (\degr) \\ \hline
443  & $44\pm12$    & 52085 & $128\pm3.8$ & 87428 & $149\pm17$  \\       
2170 & $47\pm7.6$   & 54632 & $122\pm18$  & 90659 & $144\pm2.8$ \\
8833 & $69\pm7.2$   & 57791 & $86\pm6.1$  & 91751 & $60\pm4.8$  \\
8922 & $23\pm1.8$   & 59459 & $54\pm14$   & 92512 & $91\pm11$  \\
10340 & $127\pm15$  & 59856 & $113\pm5.9$ & 92818 & $52\pm8.0$\\
10366 & $104\pm2.9$ & 61724 & $84\pm6.4$  & 93244 & $87\pm6.6$\\
10514 & $67\pm11$   & 65417 & $62\pm4.6$  & 95066 & $78\pm8.7$\\
16369 & $66\pm8.0$  & 69112 & $134\pm3.2$ & 103519& $34\pm2.5$\\
30501 & $109\pm12$  & 69879 & $90\pm9.2$  & 114421& $114\pm5.3$\\
46893 & $135\pm6.3$ & 83575 & $60\pm3.4$  &&\\
\hline
\end{tabular}
\end{table}

\begin{table}[htb]
\caption[]{\label{tab:eclipses}Tentative characteristics of the primary eclipses for the systems where $i=90\degr$ is likely.  Conjunction times are not given because their uncertainty are of the order of the orbital period.}
\begin{tabular}{lccc}\hline
HIP& Orbital Period & Duration & $\Delta m$ \\ 
& (d) & (d) & (mmag) \\ \hline
57791 & 486.7    & 4.3 & 11\\
61724 & 972.4    & 5.3 & 9\\
69879 & 212.085    & 3.7 & 9\\
92512 & 138.42    & 5.7 & 4\\
93244 & 1270.0    & 7.0 & 3\\
\hline
\end{tabular}
\end{table}
%
\subsection{DMSA/O entries}\label{Sect:DMSAO}
%
\begin{table*}[htb]
\caption[]{\label{Tab:rejDMSAO}
Statistical results \citep{Pourbaix-2001:b} for the rejected DMSA/O entries:
Pr$_1$=Pr($\hat{F_C}>F(2,N-7)$),
Pr$_2$=Pr($\hat{F}>F(4,N-9)$); Pr$_3$=Pr($\chi^2_{4S}>\chi^2(4)$); 
Pr$_4$=Pr($\chi^2_{4C}>\chi^2(4)$); $\epsilon$ is the efficiency \citep{Eichhorn-1989}; The Campbell approach may be 
accepted when the conditions Pr$_1<5\%$, Pr$_2<5\%$, 
Pr$_3<5\%$, $|D|<\approx 2$, $\epsilon\ge 0.4$, Pr$_4>5\%$, Pr$_5>5\%$, and $a_0\sin i>1.5$ mas are fulfilled.
}
\begin{tabular}{lcccccccccl}\hline
HIP &Pr$_1$ & Pr$_2$ & Pr$_3$ & $D$ & $\epsilon$ & Pr$_4$ & Pr$_5$ & Period & $a_0\sin i$ & Fitted\\
HIP &&&&&&&& (d) &(mas) & in DMSA/O\\ \hline
10324  & 15 & 21 & 18 &  -0.42 & 0.28 & 74 & 41 &  1642.1 &   7.87  & $a_0,i,\Omega$\\
13531  & 59 &  3 & 25 &  +0.47 & 0.07 & 16 & 13 &  1515.6 &  22.55  & $a_0$\\
14328  & 34 &  0 &  0 &  -3.51 & 0.33 &  0 &  0 &  5329.8 &  47.49  & $a_0$\\
17440  &  0 &  0 &  0 &  -1.98 & 0.33 &  0 &  0 &  1911.5 &  29.39  & $a_0,i,\Omega,T$\\
17932  &  0 &  0 &  0 &  +3.97 & 0.68 &  0 &  0 &   962.8 &  12.29  & $a_0,i,\Omega$\\
24608  & 99 & 53 &  1 &  -9.25 & 0.81 &  0 &  0 &   104.0 &  19.00  & $a_0$\\
24727  &  0 &  0 &  0 &  +0.25 & 0.51 &  0 &  1 &   434.8 &   7.91  & $a_0,i,\Omega,T$\\
26001  &  0 &  0 &  0 &  -1.91 & 0.80 &  5 &  1 &   180.9 &   4.47  & $a_0,i,\Omega,T$\\
29982  &  0 &  0 &  0 &  +0.17 & 0.29 & 99 & 98 &  1325.0 &  11.15  & $a_0,i,\Omega$\\
32578  &  0 &  1 &  5 &  +0.56 & 0.19 & 98 & 89 &  1760.9 &   5.13  & $a_0,i,\Omega$\\
32768  &  0 &  0 &  0 &  +5.48 & 0.75 &  0 &  0 &  1066.0 &   7.09  & $a_0,i,\Omega,T$\\
36377  &  0 &  0 &  0 &  -2.74 & 0.94 &  0 &  0 &   257.8 &   7.69  & $a_0,i,\Omega,T$\\
40326  &  0 &  0 &  0 &  -0.12 & 0.82 &  0 &  0 &   930.0 &   8.31  & $a_0,i,\Omega,T$\\
45527  & 99 &  0 &  0 &  +5.88 & 0.63 &  0 &  0 &   922.0 &   8.84  & $a_0,i,\Omega,T$\\
49841  &  0 &  0 &  0 &  -0.20 & 0.24 &  0 &  0 &  1585.8 &  15.18  & $a_0,i,\Omega,T$\\
53240  &  0 &  0 &  0 &  -2.00 & 0.46 &  4 &  0 &  1166.0 &   5.66  & $a_0,i,\Omega$\\
57565  & 27 & 39 & 50 &  -0.87 & 0.68 & 91 & 65 &    71.7 &   2.78  & $a_0$\\
63613  & 99 &  0 &  0 & -12.66 & 0.73 &  0 &  0 &   847.0 &  19.17  & $a_0,i,\omega,\Omega,e,P,T$\\
67234  &  0 &  0 &  0 &  +1.02 & 0.84 &  0 &  0 &   437.0 &   6.31  & $a_0,i,\Omega,T$\\
80166  &  1 &  0 &  5 &  -0.86 & 0.46 & 83 & 57 &   922.8 &   1.91  & $a_0,i,\Omega$\\
80816  &  0 &  0 &  0 &  -0.24 & 0.57 &  0 &  0 &   410.6 &   8.65  & $a_0,i,\Omega,T$\\
85749  &  0 &  0 &  0 &  +0.59 & 0.61 &  1 & 12 &   418.2 &   5.81  & $a_0,i,\Omega$\\
96683  & 99 &  0 &  5 &  -0.75 & 0.71 &  0 &  0 &   434.1 &  11.59  & $a_0$\\
110130 &  0 &  0 &  0 &  -0.82 & 0.13 &  0 &  0 &  4197.7 &  42.38  & $a_0,T$\\
112158 &  0 &  0 &  0 &  -3.81 & 0.81 &  0 &  0 &   818.0 &  15.55  & $a_0,i,\Omega,T$\\
\hline
\end{tabular}
\end{table*}

As already stated, a lot of the DMSA/O entries (i.e. objects whose observations were initially fitted with an orbital model) get rejected although the inclinations derived with the two methods agree very well (right panel of Fig.~\ref{fig:inclinations}).  Table \ref{tab:inclinations} is therefore likely to be (too) conservative and some further investigation of the DMSA/O entries seems worth undertaking.

The 25 DMSA/O entries rejected (Table~\ref{Tab:rejDMSAO}) can be subdivided into cases where the spectroscopic constraints have to be slightly relaxed on the one hand and those where there is no noticeable and/or reliable improvement with the orbital model on the other hand.  For about 50\% of these systems, $T$ (the periastron time), though known a priori, was adjusted also.  For none of the latter, the orbit is circular.  One can therefore wonder how reliable this approach is.  Indeed, even if $T$ is not well estimated from the spectroscopic orbit, neither does $\omega$ (the argument of the periastron).  It therefore does not make much sense to change $T$ without also fitting $\omega$.

\begin{figure*}[htb]
\resizebox{0.25\hsize}{!}{\includegraphics{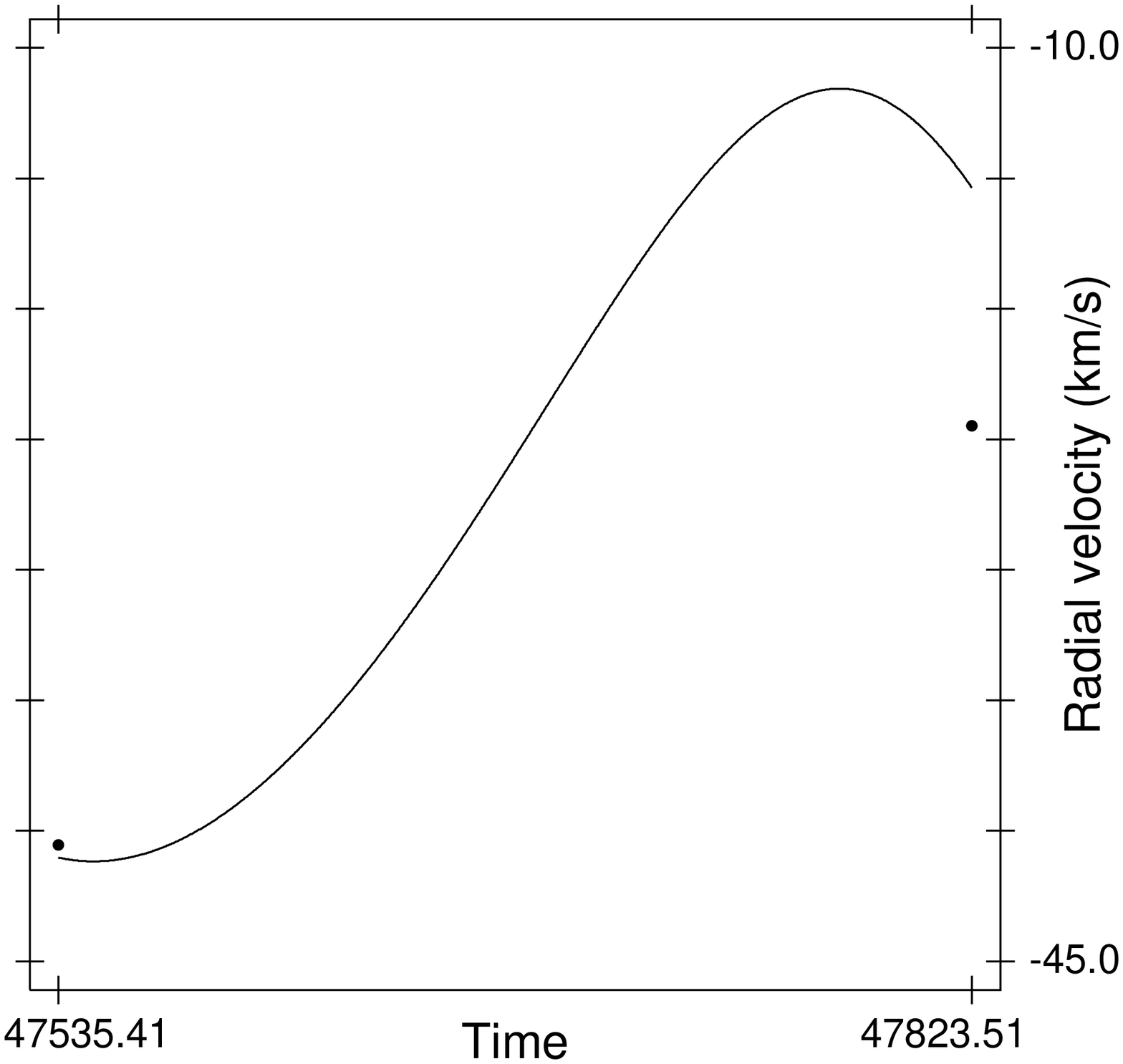}}\hfill
\resizebox{0.25\hsize}{!}{\includegraphics{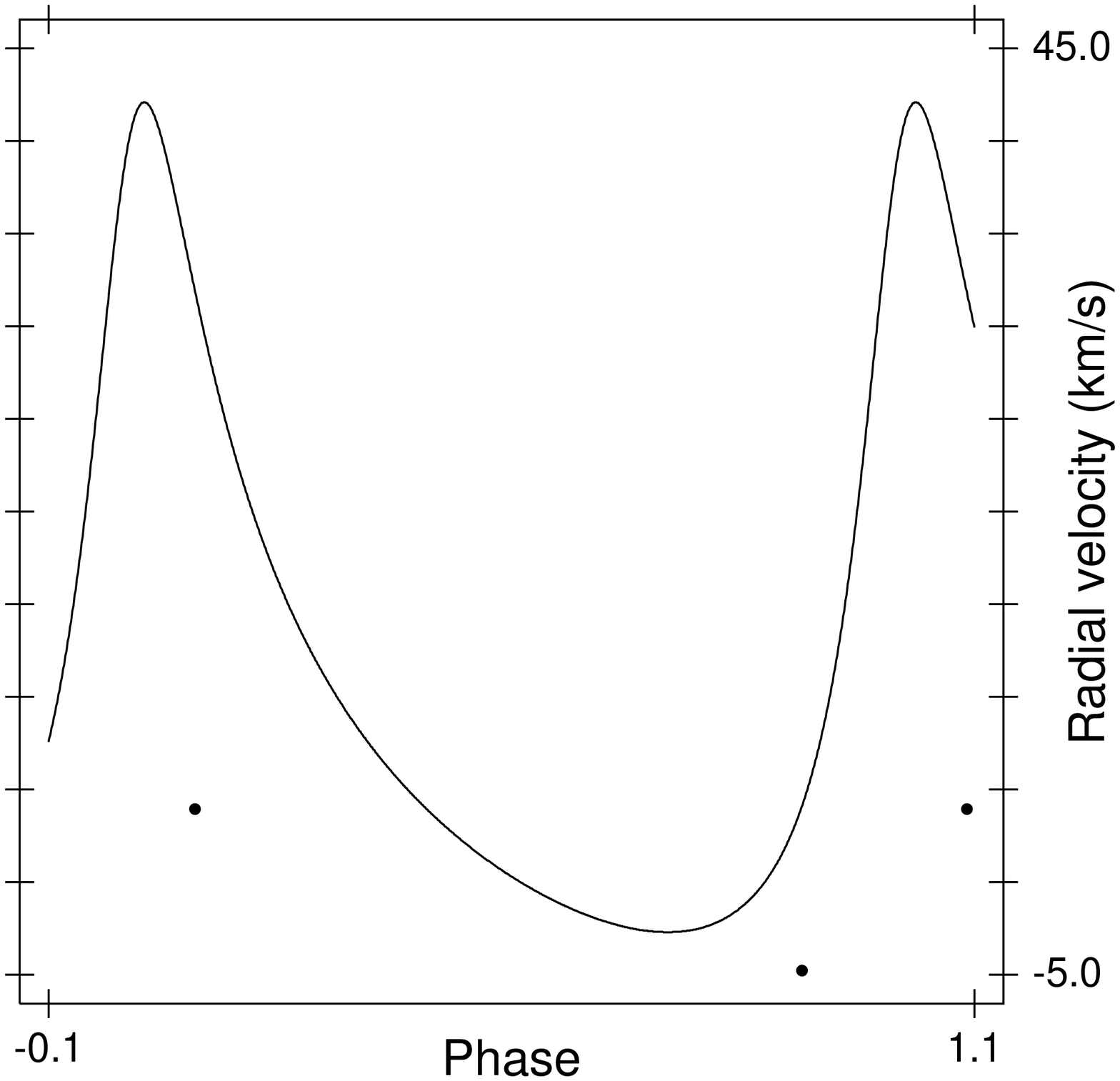}}\hfill
\resizebox{0.25\hsize}{!}{\includegraphics{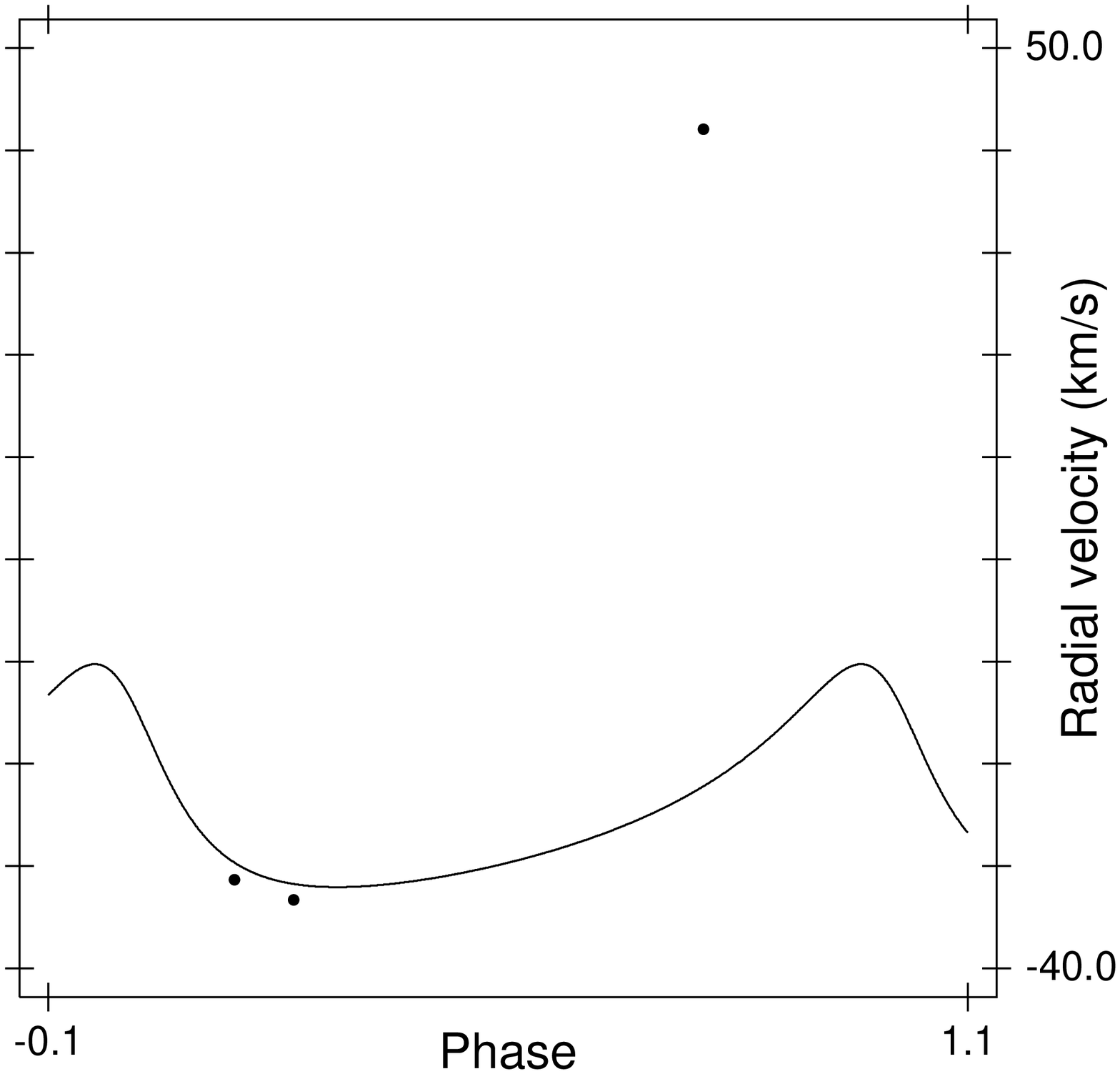}}\hfill
\resizebox{0.25\hsize}{!}{\includegraphics{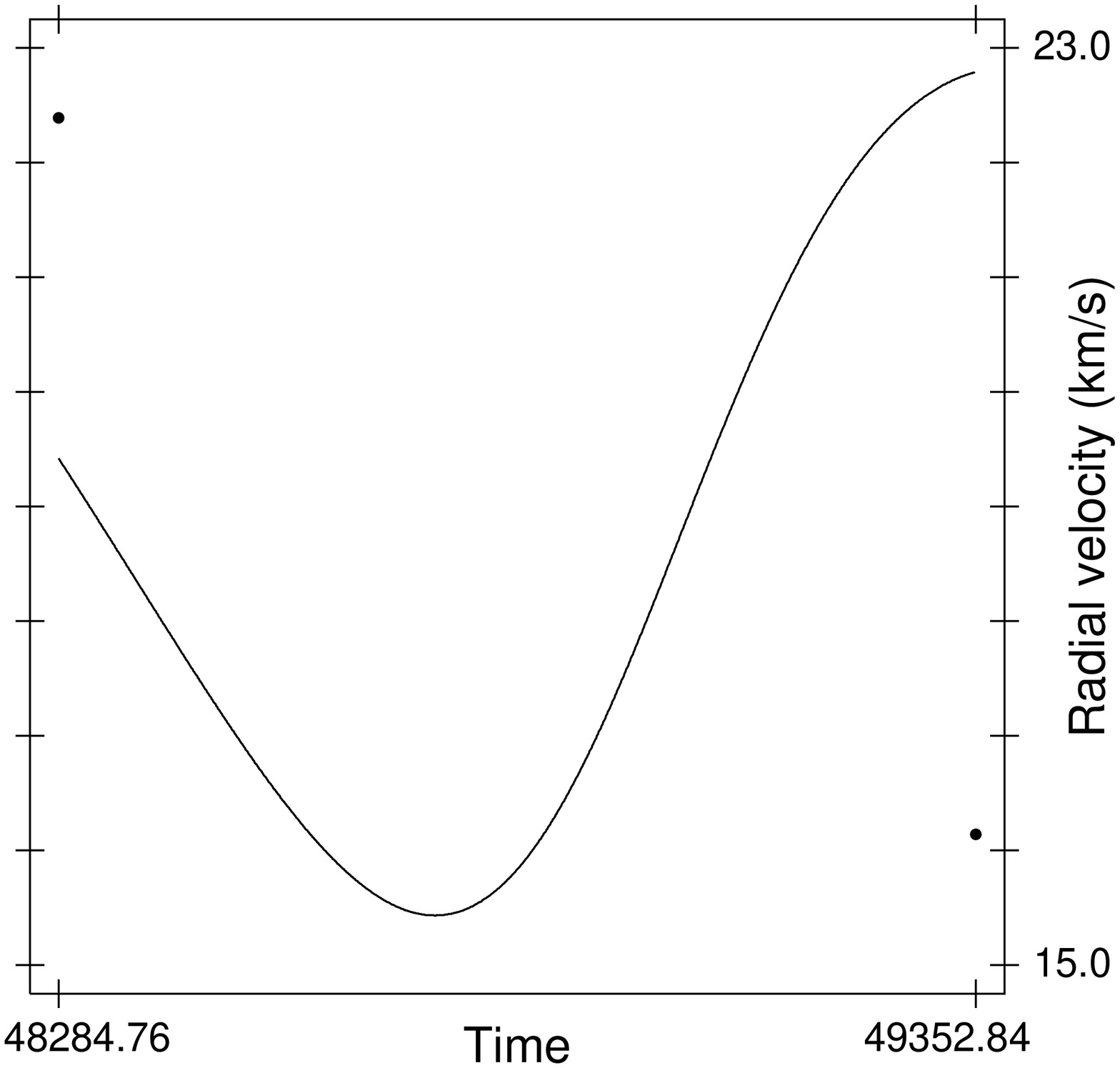}}\\
\resizebox{0.25\hsize}{!}{\includegraphics{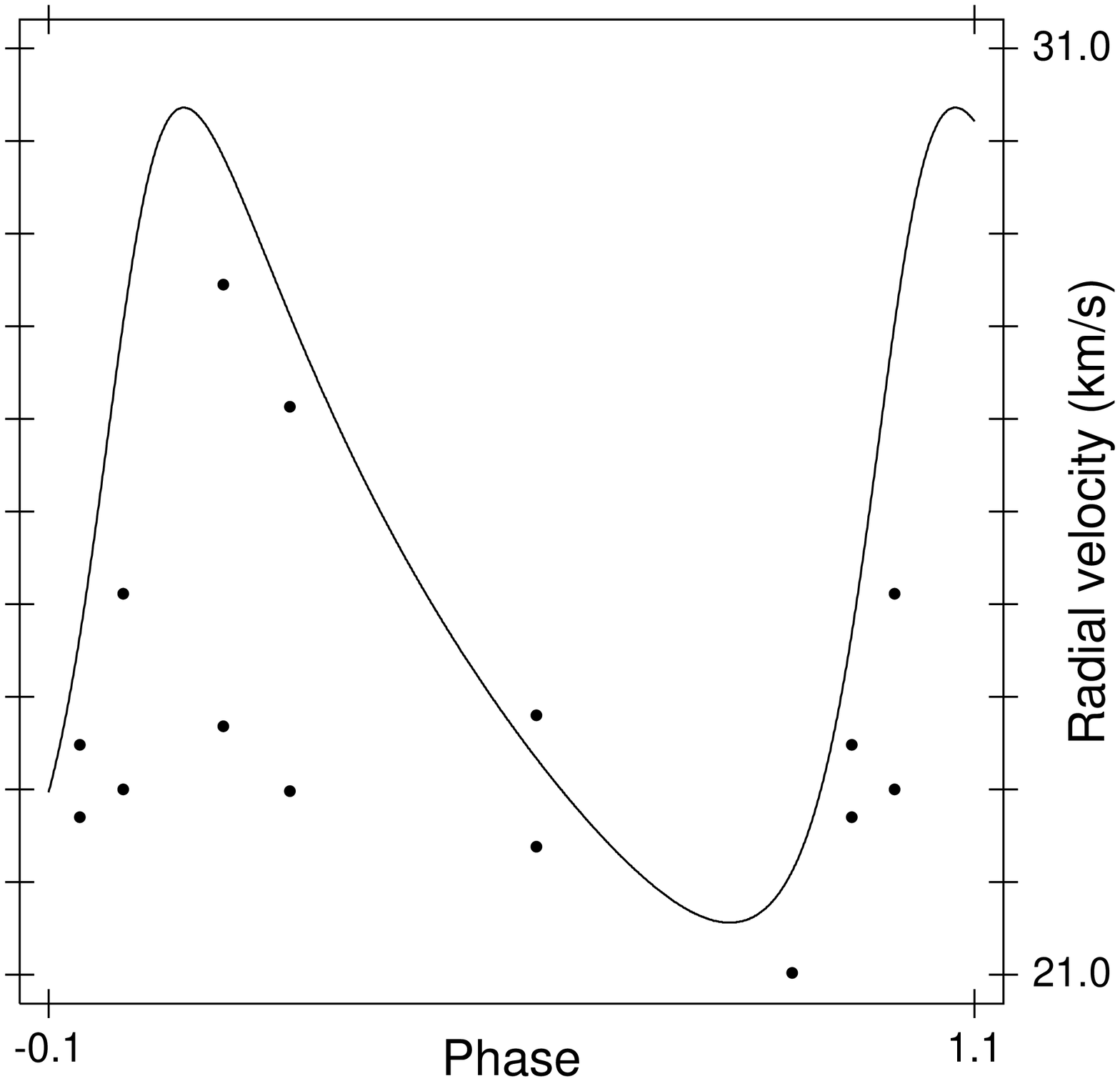}}\hfill
\resizebox{0.25\hsize}{!}{\includegraphics{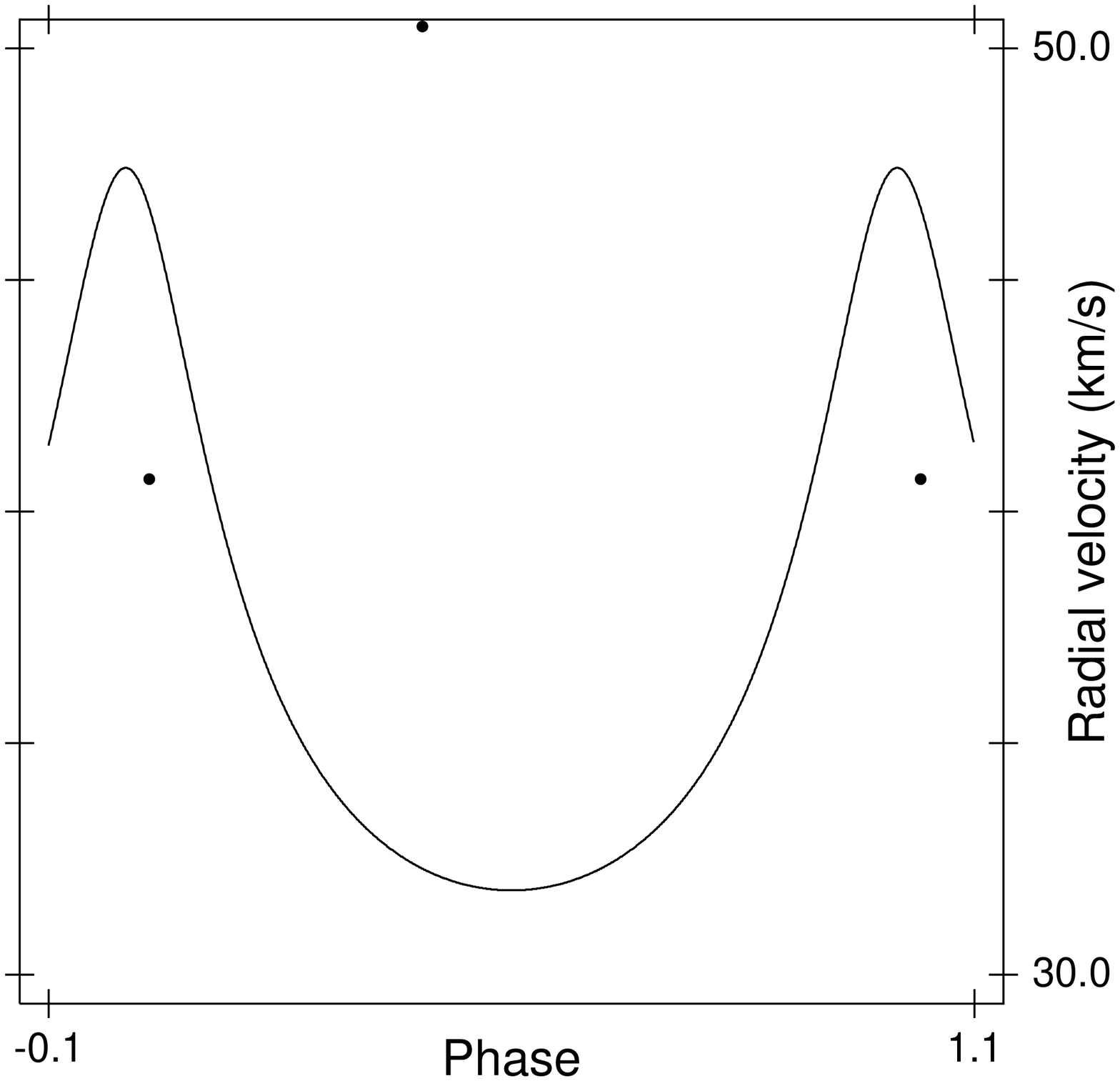}}\hfill
\resizebox{0.25\hsize}{!}{\includegraphics{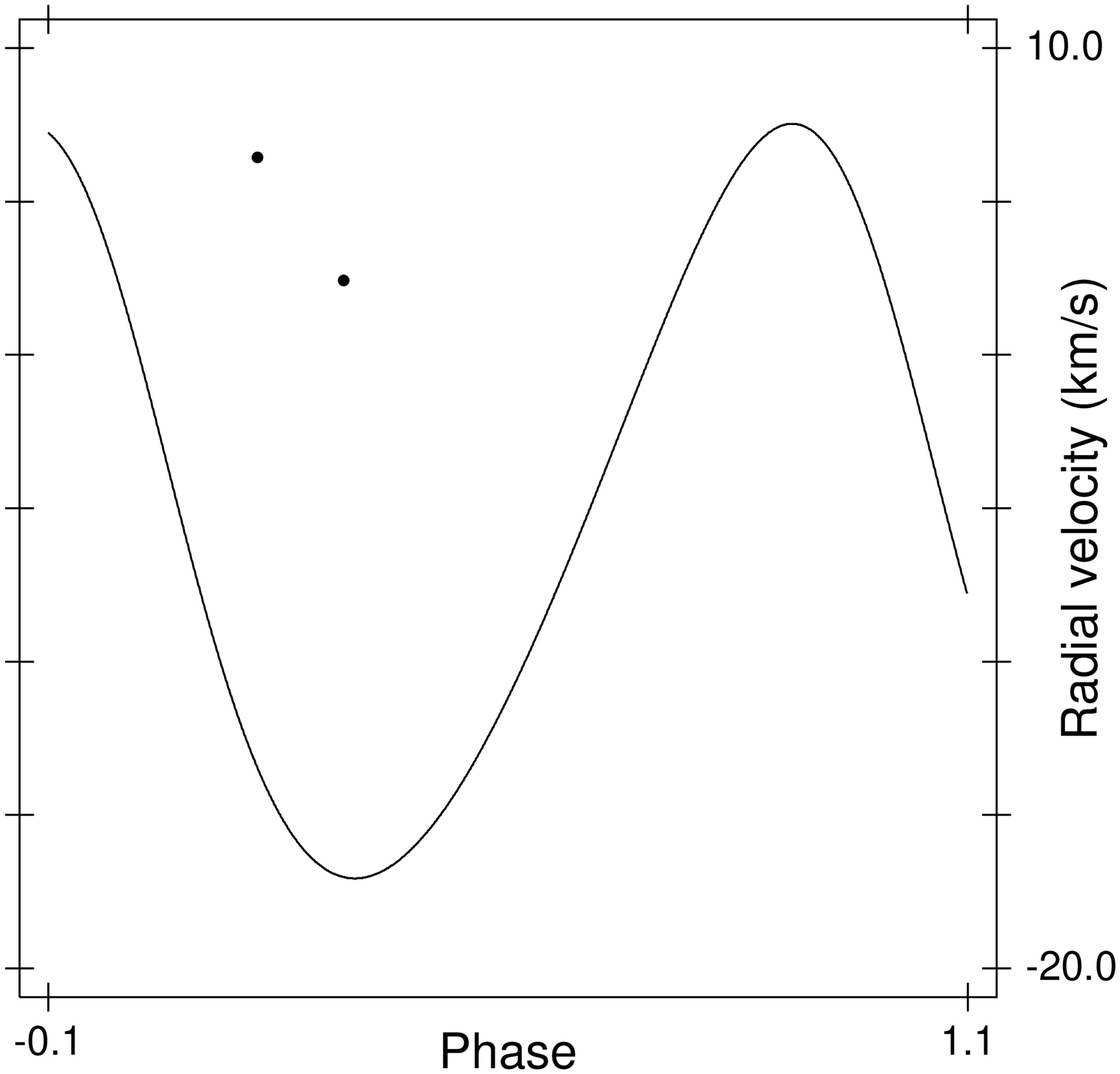}}\hfill
\resizebox{0.25\hsize}{!}{\includegraphics{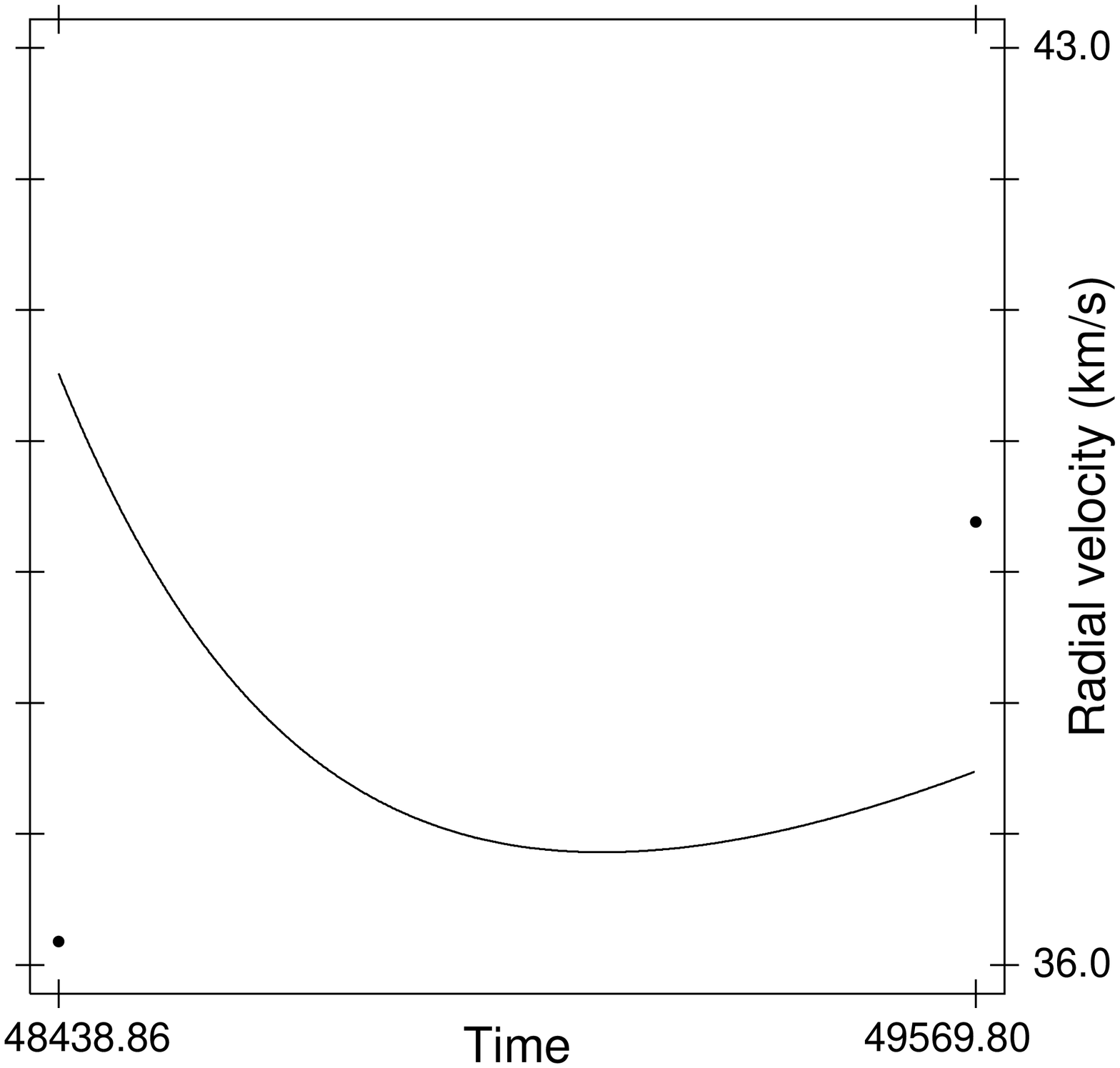}}
\caption[]{\label{fig:CORAVEL}Comparison of the adopted orbits with the CORAVEL data and the quality flag from \citet{Batten-1989:a}, ranging from e: `very poor and unreliable orbit' to b: `good orbit, but not definitive'.  From left to right, the systems are: HIP~24727 (d), 26001 (b), 29982 (b), 49841 (c) (upper panel), 53240 (c), 63613 (e), 67234 (c), and 110130 (c) (lower panel).}
\end{figure*}

For 19 systems among the rejected ones, either $Pr_4$ or $Pr_5$ (but usually both) is very close to 0 although the periodogram exhibits a peak at the expected period.  $Pr_4\approx 0$ indicates that, even if both Campbell's (14 cases) and Thiele-Innes' (18 cases) solutions do improve the fit, they do not agree with each other.  In order to assess the reliability of the spectroscopic orbit assumed in the fit, that orbit is tested against the CORAVEL radial velocities (S.~Udry, private communication) obtained in the framework of the measurement of the whole Hipparcos catalogue with CORAVEL (Fig.~\ref{fig:CORAVEL}).  Some of the adopted orbits are obviously disproved by those modern radial velocities.

%
\subsection{Additional entries}\label{Sect:NewOrb}
%

There are six stars for which the orbital model seems appropriate even though they do not belong to the DMSA/O: HIP 2170 (DMSA/C), 8922 (DMSA/X), 10340, 46893 (DMSA/X), 54632, and 87428.  The astrometric orbits are given in Table \ref{Tab:NewOrb}.  

\begin{table*}[bth]
\caption[]{\label{Tab:NewOrb}Six new orbital solutions}
\begin{tabular}{lcccccc}\hline
HIP                & 2170       & 8922         & 10340        & 46893       & 54632      & 87428 \\ \hline
$a_0$              & $6\pm2.0$  & $8.1\pm0.93$ & $3.0\pm2.6$  & $4.8\pm0.8$ &$2\pm0.55$ & $5.3\pm1.2$\\
$i$ (\degr)        & $47\pm7.6$ & $24\pm1.8$   & $127\pm15.0$ & $135\pm6.3$ &$122\pm18$ & $149\pm17$\\
$\omega_1$ (\degr)   & --         & --           & $358\pm6$    & $261\pm12$  & $332.97\pm4.23$        & --\\
$\Omega$ (\degr)   & $300\pm10$ & $155\pm4.0$  & $6\pm17$     & $5\pm11$    &$254\pm22$  & $163\pm15$\\
$e$                & 0          & 0            & $0.34\pm0.03$  &$0.149\pm0.030$& $0.282\pm0.017$       & 0\\
$P$ (d)            & $936\pm4$  & $838\pm4$    & $748.2\pm0.4$&$830.4\pm2.5$& $18.8922\pm0.0061$    & $467.2\pm1.2$\\
$T$ (HJD) (2.4E6+) &$42531.4\pm2.6$& $43521\pm5$  & $37886\pm11$ & $43119\pm25$& $23154.071\pm0.178$  & $42484.9\pm2.7$\\
Ref                & \citet{Griffin-1979:a} & \citet{Griffin-1981:b} & \citet{Griffin-1981} & \citet{Griffin-1981:c} & \citet{Sanford-1924:a} & \citet{Griffin-1980:b}\\
\hline
\end{tabular}
\end{table*}

The case of HIP~2170 is worth mentioning.  On the one hand, the companion responsible for the DMSA/C entry was first ever detected by Hipparcos and no ground-based observation has so far confirmed the presence of that companion.  On the other hand, the periodogram does exhibit its deepest peak at the spectroscopic orbital period.  So it is likely that what Hipparcos saw is actually that spectroscopic companion.  With the orbital solution instead of the DMSA/C one, the parallax changes from $1.61\pm1.24$ mas to $3.65\pm0.94$ mas, yielding $M_V\sim 1$ (instead of -0.76).

%
\subsection{Biases}\label{Sect:biases}
%

\begin{figure}[htb]
\resizebox{\hsize}{!}{\includegraphics{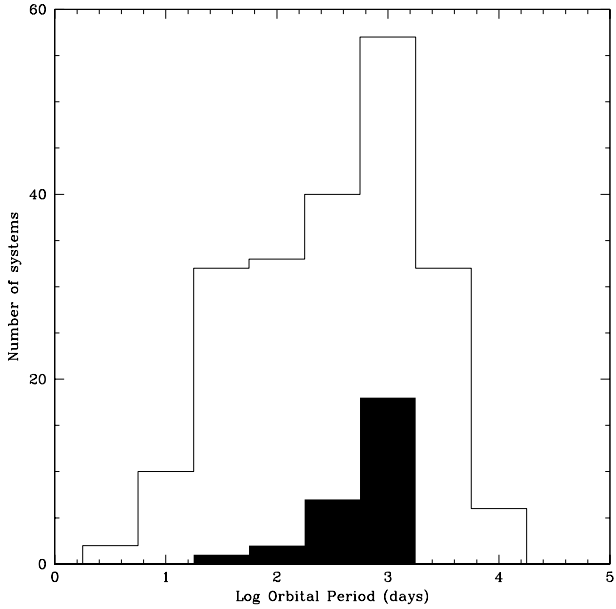}}
\caption[]{\label{fig:porb} Distribution of the logarithm of the orbital
period for the full sample and for our selected subsample of 29 systems (shaded
area). The subsample shows a strong peak for systems between 400 and 1400 days,
}
\end{figure}

The BCP93 sample of spectroscopic binaries containing a red giant 
and its present extension clearly suffer from many observational
biases: it is by no means complete in magnitude, it is restricted globally 
to systems with large enough radial velocity semi-amplitude (i.e. small mass ratios
and/or long periods are underrepresented), restricted by definition in the mass range of the primary
but not necessarily in the binary population, age, etc.
Trying to correct for these biases is a rather challenging task, well beyond
the scope of this paper, and one should thus be aware that the results we
derive should not be taken as such for the study of binary star formation
for example. 
Already the fact that we are looking only at red giant primaries
implies that one should correct everything for the stellar evolution effects.
However, as was already the case for BCP93, the idea here is not to obtain
directly useful hints on the way stars do form but instead to study the 
properties of a sample which can be used as a comparison for Peculiar Red Giant 
stars, and hence which suffers from similar biases. 
The subsample of 29 orbits that we have selected will clearly suffer at least 
from the same biases. We have however checked that except for the orbital period, 
there is statistically no difference between our subsample and the full sample of
215 stars, regarding for example the visual magnitude, mass function or eccentricity distributions. 
As far as the orbital period is concerned, our subsample is strongly peaked
(see Fig.~\ref{fig:porb}) in the period interval between 400 and 1400 days, 
with the smallest period being about 19 days and the longest 1672 days. 
There is no obvious reason however why this selection in period should impose
a bias on the masses of the two components (but see below) and we may therefore believe that our subsequent analysis is representative of the whole sample of red giant SBs.
One should note that the system with the shortest period (19 d) has in fact a 
main-sequence primary (see below).

%
\section{Mass function and masses of the primary}\label{Sect:massfuncprim}
%
With the knowledge of the inclination as given by Table~\ref{tab:inclinations}, we can estimate for each of the 29 systems, from the mass function $f(m)$, 
the quantity denoted by $\cal{Q}$ 
in BCP93, and which is a function of the masses of the two components ($m_1$ \& $m_2$) of the
binary system:
\begin{equation}
{\cal Q} = \frac{f(m)}{\sin ^3 i} = \frac{m_2^3}{(m_1+m_2)^2}.
\end{equation}

\begin{figure}[htb]
\resizebox{\hsize}{!}{\includegraphics{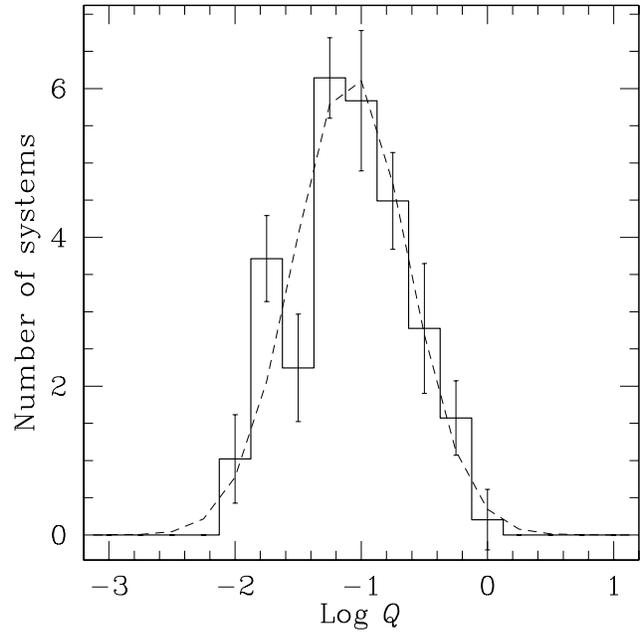}}
\caption[]{\label{fig:Qdis}
Distribution of the logarithm of $\cal{Q}$ for the sample of
28 single-lined systems which we retained.
The error bars correspond to a Monte Carlo estimate taking into
account the errors on $\cal{Q}$. The dashed curve is a Gaussian
fit to the data, with a mean at -1.08 and a half width at half maximum of 0.45.  }
\end{figure}

The resulting distribution of the logarithm of this quantity is shown in Fig.~\ref{fig:Qdis}.
It can be seen that the distribution is compatible with a Gaussian 
with a mean at -1.08 and a sigma of 0.45.
Such a mean value corresponds to a mass of a companion of 0.7, 0.9 and 1.1 M$_\odot$
for a primary mass of 1.5, 2 and 3 M$_\odot$, respectively. 
There is a small excess of systems having $\log$~$\cal{Q}$ around -1.75.
This value of $\cal{Q}$ corresponds to a mass of a companion of 0.4, 0.5
and 0.6 M$_\odot$ for the above quoted value of the primary mass.

We have also performed a comparison between the distribution of $\cal{Q}$
one would obtain assuming a random inclination, both for our subsample of 28
systems and for the full sample of 215 systems. A Kolmogorov-Smirnov test 
could not distinguish between the two derived distributions. This confirm
that our subsample should be statistically representative of the full 
sample of red giants spectroscopic binaries.

In the case of single-lined spectroscopic binaries, and even when one knows the inclination of the system, there is no immediate way to determine the mass of the components, not even for the primary. In our case, as the systems have red giant primaries, it is also not possible to make use of a simple mass-luminosity relation to estimate the mass as would be the case - as a first approach - for main-sequence stars.  If sufficiently accurate information is available, however, on the effective temperature - either through spectroscopy or from the colors - and on the bolometric luminosity - through a good knowledge of their parallax - one can hope to estimate a reasonable mass of the primary by using an Hertzsprung-Russel diagram and compare with stellar evolutionary tracks.  As the 29 stars we have selected were, by definition, observed by Hipparcos and reanalyzed by us, a good estimate of the parallax and its associated error is available.  The stars being bright, good values of their magnitude and $(B-V)$ color index are available.  Moreover, \citet{McWilliam-1990:a} has produced a high-resolution spectroscopic survey of 671 G and K field giants which includes most of them.  This allows to have a good determination of the effective temperature as well as of the metallicity of our stars.  Most of our stars have solar metallicity, with the most metal-poor star being HIP~90659 with a [Fe/H]=-0.67.

We have thus obtained for the 29 stars in our significant sample the bolometric magnitude ($M_{\rm bol}$) and the effective temperature ($T_{\rm eff}$).  For $T_{\rm eff}$, we searched for values in the literature, with most values being obtained from \citet{McWilliam-1990:a}. If no value was found, we estimated one from the $(B-V)$ index using the relation  \citep{McWilliam-1990:a} : 
\begin{equation}
T_{\rm eff} = 8351 - 4936~(B-V) + 1456~(B-V)^2 - 78~(B-V)^3.
\end{equation}

The bolometric magnitude was obtained from the parallax ($\varpi$, in mas) and bolometric correction ($BC$), 
through the relations :
\begin{equation}
M_{\rm bol} = m_V - 5~log \frac{1000}{\varpi} + 5 - A_V + BC,
\end{equation}
where $A_V$ is the visual extinction given by version 2.0.5 of the EXTINCT subroutine \citep{Hakkila-1997:a}.
The value of $BC$ was estimated from $T_{\rm eff}$ using the tables of \citet{AsDaPlSt}.

\begin{figure*}[htb]
\resizebox{0.49\hsize}{!}{\includegraphics{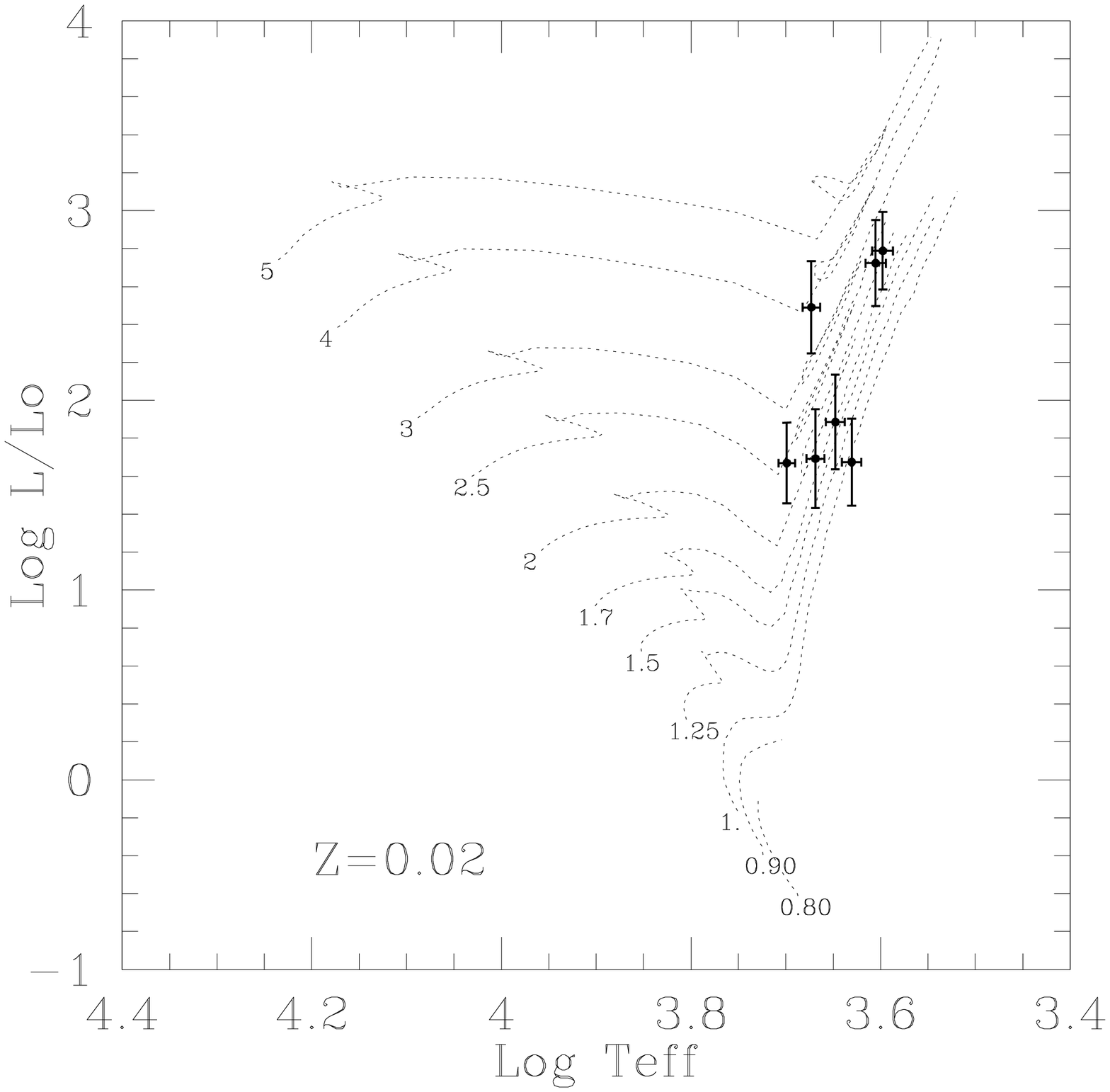}}
\resizebox{0.49\hsize}{!}{\includegraphics{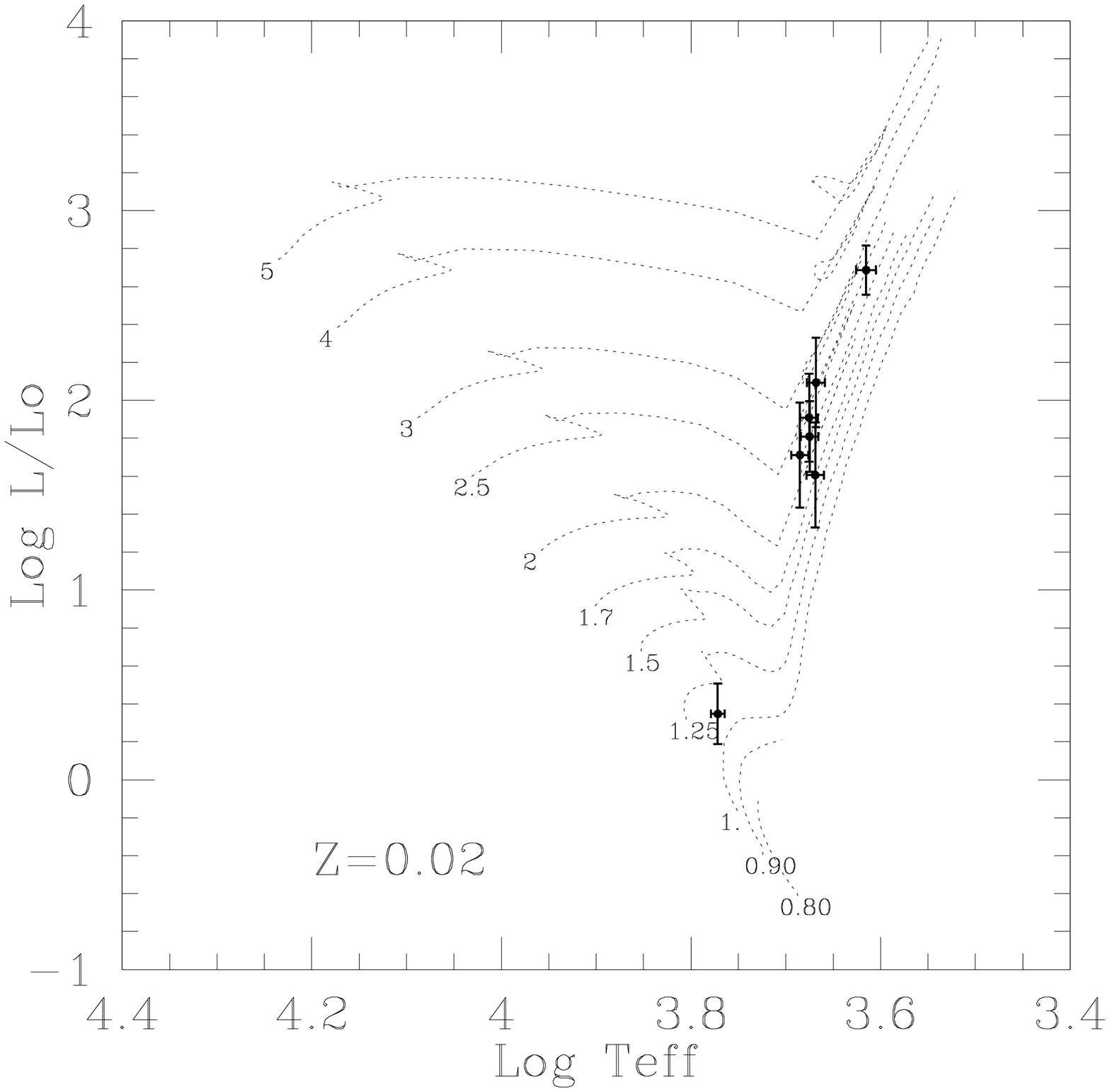}}\\
\resizebox{0.49\hsize}{!}{\includegraphics{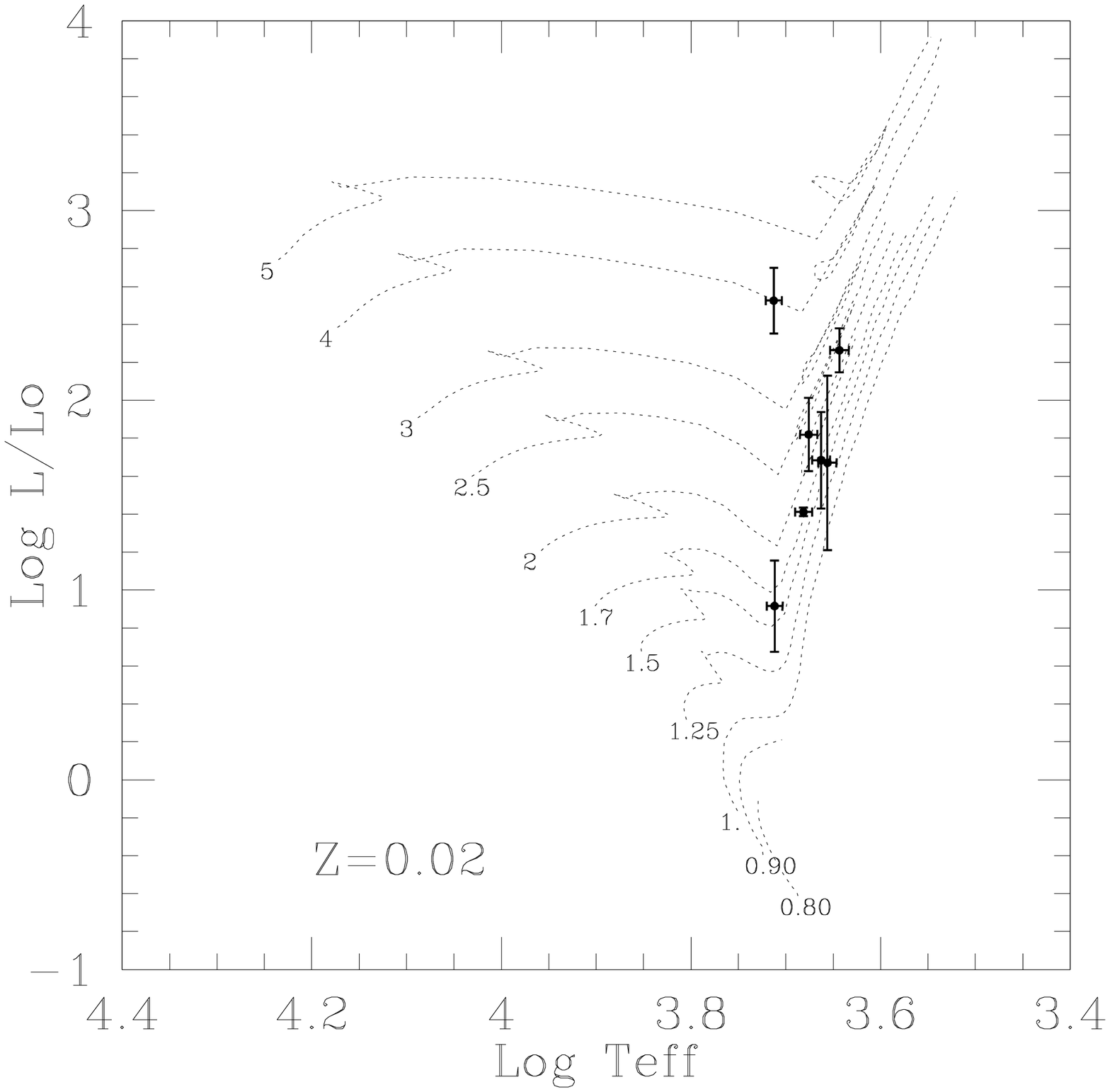}}
\resizebox{0.49\hsize}{!}{\includegraphics{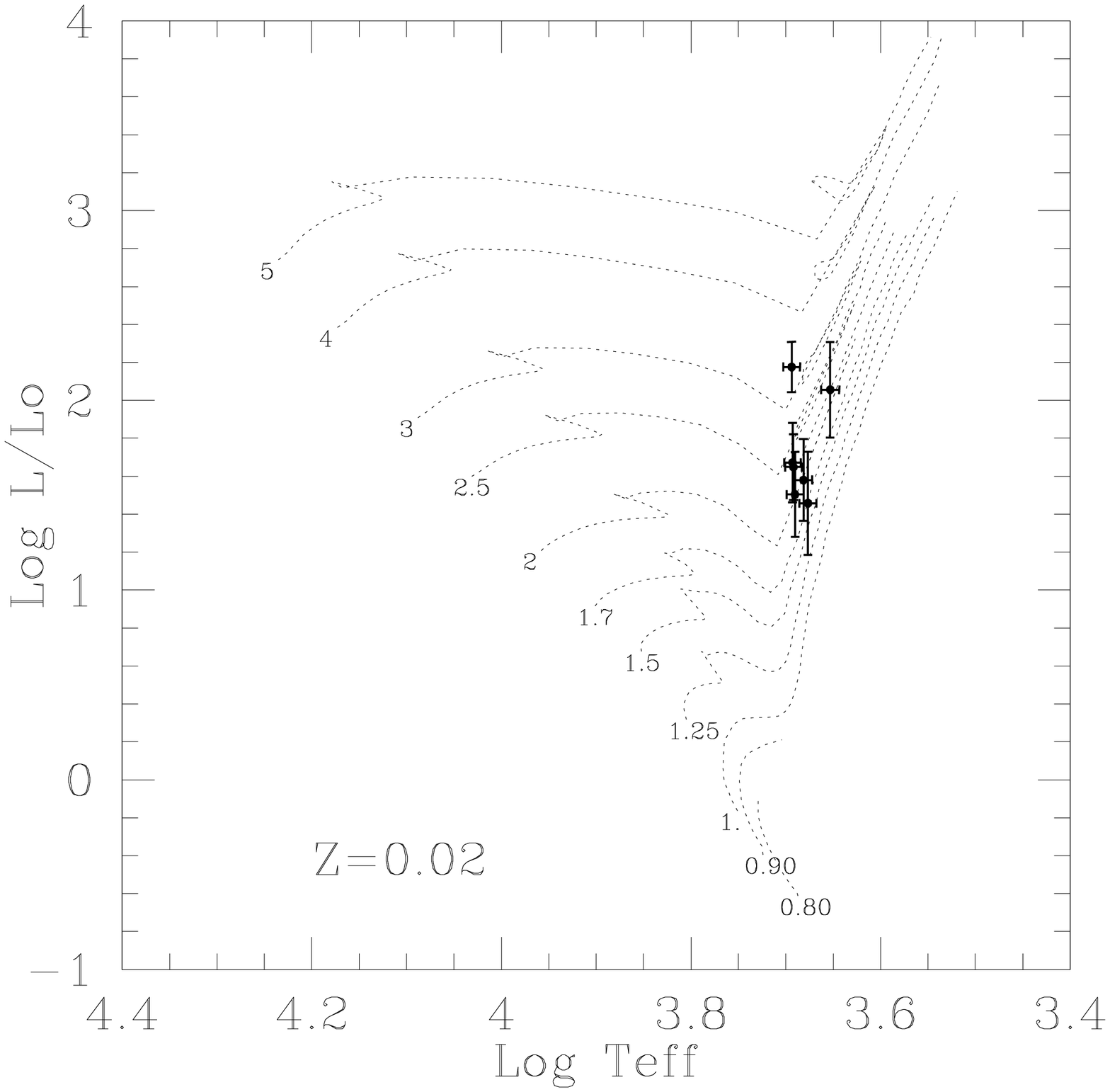}}
\caption[]{\label{fig:HRdiag}
Position of the retained 28 single-lined systems in the H-R diagram.
For comparison, we show the stellar evolutionary tracks for 
Z=0.02 stars as obtained by the Geneva group. The curves are labeled with the initial
stellar mass on the main sequence, in solar masses.
}
\end{figure*}

The results are shown in Fig.~\ref{fig:HRdiag}, where for clarity we have separated our sample into 
4 arbitrary sub-samples. 
One sigma errors are also indicated. 
On these figures, we also show the evolutionary tracks for stars of
solar metallicity with masses between 0.8 and 5 M$_\odot$ as obtained by the Geneva group \citep{Schaller-1992,Charbonel-1996:a}.
From the comparison of the positions of these stars and their uncertainties, one can estimate
the probable range of mass of the primary of each system.
One has to note that because these systems are single-lined spectroscopic binaries, the secondary 
should not contribute too much light to the whole system, Hence, using the total luminosity of the 
system and color of the system to derive the mass of the primary should not be a large source of error. 
This will be checked a posteriori (see Sect.~\ref{Sect:masssec}).
As all the systems for which the metallicity is known have close to solar abundance, we do not need
to compare our data with stellar evolutionary tracks corresponding to other value for Z than 0.02.
All our results are summarized in Table~\ref{tab:results} where we indicate, respectively, the HIP 
identification number, the effective temperature, the parallax and its sigma deviation, the bolometric
magnitude, the mass of the primary, the lower and upper 
limits on the mass of the primary, the mass of the secondary, and the mass ratio. 

\begin{figure}[htbp]
\resizebox{\hsize}{!}{\includegraphics{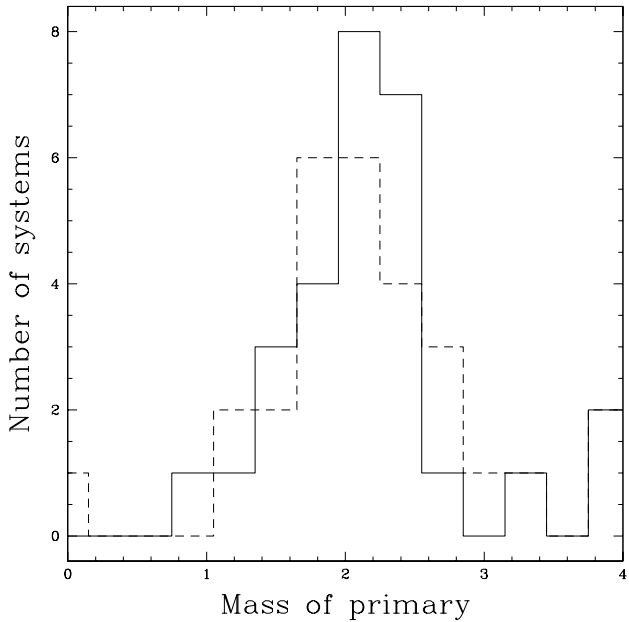}}
\caption[]{\label{fig:massdis} Distribution of the masses of the primary for 
the 28 single-lined SBs. The solid line shows the distribution taking the face
value of the primary masses, while the dashed line shows the distribution 
with the possible range for each primary mass as derived from the one sigma errors shown 
in the HR diagram.
}
\end{figure}

The distribution of primary masses are shown in Fig.~\ref{fig:massdis}.
It is strongly peaked around 2 M$_\odot$ with only 3 out of 28 systems having a primary 
mass equal or above 3 M$_\odot$. 
A value of 2 M$_\odot$ is consistent with the results of \citet{Strassmeier-1988:a} who obtain the
same value for the red giant component of binary systems. More recently, \citet{Zhao-2001:a} also 
concluded that the majority of red clump giants have masses around 2 M$_\odot$. 
BCP93 favored a mass of the red giant primary of 1.5 M$_\odot$
in their statistical analysis, while \citet{Trimble-1990:a} assumed a mass of 3 M$_\odot$. From our
analysis, this last value seems too large, while the first appears slightly too small.

%
\section{Masses of secondary and mass ratio distribution}\label{Sect:masssec}
%
With the value of the primary mass deduced from the H-R diagram and the value of
the quantity $\cal{Q}$ deduced from the inclination and the mass function, one can now
determine for each system the mass ratio and the secondary mass. This is a 
straightforward exercise which only requires to solve a third degree polynomial.
The values for the individual masses will be discussed later (see Sect.~\ref{Sect:individual}).
Here we will look at the distribution of these quantities so as to compare with the
previous analysis of BCP93 and others.

\begin{figure}[htb]
\resizebox{\hsize}{!}{\includegraphics{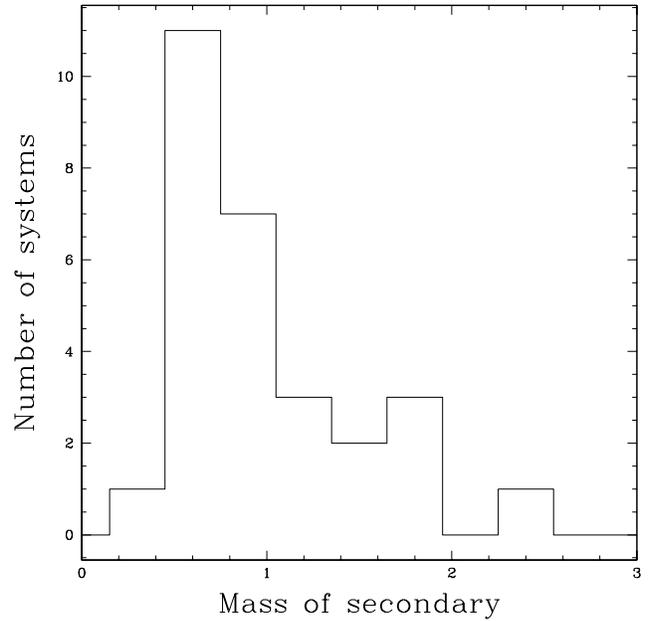}}
\caption[]{\label{fig:masssec} Distribution of the masses of the secondary for
the 28 single-lined SBs.}
\end{figure}

\begin{figure}[htb]
\resizebox{\hsize}{!}{\includegraphics{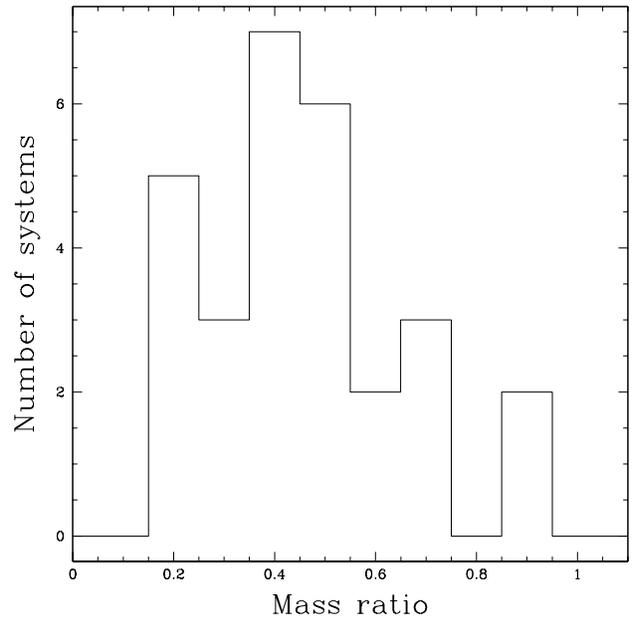}}
\caption[]{\label{fig:massrat} Distribution of the mass ratio for
the 28 single-lined SBs.}
\end{figure}

The distribution of the secondary mass ($m_2$) is shown in Fig.~\ref{fig:masssec}.
It appears to be exponentially increasing for smaller masses with a drop-off
for secondary masses below 0.5 M$_\odot$. This drop-off must be related to the absence
of systems in our sample with a mass ratio below 0.2 (see below).
We have checked that the distribution of $m_2$ can be fitted with a Salpeter-like IMF with $\alpha=2.35$ when for each system $m_2$ is restricted in the range
from 0.55 M$_\odot$ to $m_1$. The upper limit is straighforward as we are 
dealing with red giant primaries and the companion must be either less evolved 
and hence less massive, or must be a white dwarf and therefore 
less massive than the giant whose mass is generally above 1.4 M$_\odot$. The
lower limit can be derived from the fact that our subsample contains mostly
systems for which the orbital period is larger than 400 days, the radial velocity semi-amplitude is larger than 10 kms$^{-1}$ and the mass of the primary
is larger than  1.4 M$_\odot$. From the definition of the mass function, 
this provides a lower limit to the mass of the secondary. This is a clear 
observational bias against low mass secondaries but should not question the
validity of the apparent IMF distribution of secondary masses.

The median of the distribution is around 0.8 M$_\odot$ with the largest peak around 0.75 M$_\odot$, 
even though there is also a small clustering around 0.52 M$_\odot$. 
There are 4 systems - excluding the double-lined spectroscopic binary (see below) - with 
a secondary mass above 1.5 M$_\odot$. One should be careful however that these systems might be
masqueraders, as in the largest systems, the secondary might well be itself a close binary.
This is a scenario which seems very probable for HIP 65417 which has a secondary mass of
2.5 M$_\odot$ for a primary mass of 2.7 M$_\odot$ in a long period system, 1367 days. In fact, 
given these values, this systems is the one with the largest semi-major axis, 4 AU. In such 
a system, it is no problem to have a hierarchical triple system with the secondary being 
itself a close binary system composed of two F or G main sequence stars with an orbital 
period of a few tens of days. This was in fact already suggested by \citet{Griffin-1986:b}.

For the other systems, we have checked that there is no apparent correlation between the
mass of the secondary and the orbital period (or the semi-major axis) so that there is 
no reason to believe in having a triple system.

We now turn to the distribution of the mass ratio which is shown in Fig.~\ref{fig:massrat}.
Two facts are easily seen. First, the distribution clearly increases for smaller mass ratios.
Second, there are no system with a mass ratio below 0.2. 
This second effect is clearly a bias related to the need to find an astrometric signal for the 
stars we have selected. Such a signal becomes weaker for smaller mass ratios, as in the case
where the secondary does not contribute to the light of the system, the motion of the
photocenter is given by the motion of the primary, hence:
\begin{equation}
a_0 \equiv a_1 = \frac{m_2}{m_1+m_2} a  = \frac{q}{1+q} a. 
\end{equation}
For the most close (on the sky) systems, it is not possible to detect the motion if the mass
ratio is too small. And indeed, except for one case, there are no systems with a mass ratio below
0.4 with an angular semi-major axis smaller than 15 mas. 

The distribution of mass ratios increasing for smaller mass ratios is thus even stronger than 
we obtain. One should note that there is no obvious reason why there should be a bias against 
larger mass ratios. Indeed, except for the case of $q$ very close to 1, resulting in a system with 
two giants, hence a double-lined binary, a larger $q$ makes the astrometric signature more visible. 
And from the full catalogue, the number of SB2 is rather limited (see also BCP93). 

BCP93 had to assume a constant mass for the red giant primary - or an IMF-type
distribution - and found that the mass ratio distribution (MRD) is
either compatible with a uniform distribution for  $m_1=1.5$ M$_\odot$
but more peaked towards smaller mass ratio for $m_1=3$ M$_\odot$.
\citet{Heacox-1995:a} performed an analysis of the same sample as BCP93 and found a
MRD peaked at values of the mass ratio between 0.3 and 0.5, with a drop off
at smaller values. 
Using a subsample of K giant primaries, \citet{Trimble-1990:a} derived a 
$q^{-1}$ MRD but BCP93 have shown that this is in fact an artifact of the
method used and does not reflect the true MRD of the sample.
The MRD we derive here seems to be even more peaked than $q^{-1}$. 
To better understand this, we have also derived the MRD from our 28 systems, 
assuming - as BCP93 and \citet{Heacox-1995:a} - a single value of 1.5  M$_\odot$ for
$m_1$. In this case, we obtain a similar MRD as the one obtained by Heacox.
The fact that the MRD we obtain here is even more peaked in the range 0.3-0.5 
must be related to the non-uniqueness of the mass of the primary and the fact 
that it is more strongly peaked at 2 M$_\odot$ instead of 1.5 M$_\odot$. 

\begin{figure}[htb]
\resizebox{\hsize}{!}{\includegraphics{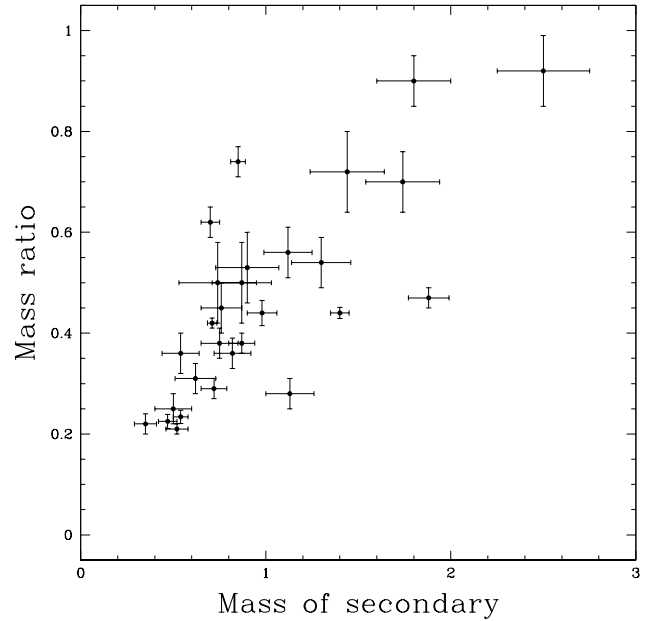}}
\caption[]{\label{fig:mrm2co} Mass ratio vs. the secondary's mass for 
the 28 single-lined SBs. A clear trend is visible.}
\end{figure}

We have also tried to see if there was any correlation between the masses of the two components
and/or with the mass ratio. There is clearly no correlation between the mass of the primary and
the mass ratio. Nor is there any significant relation between the masses of the two components.
A clear trend is however seen between the mass of the secondary and the mass ratio. This is 
illustrated in Fig.~\ref{fig:mrm2co}.
This might be a consequence of the strongly peaked value for the primary mass. 

\begin{table*}
\caption{\label{tab:results} Parameters of the 28 single-lined systems we retain}
\begin{tabular}{lllllllll}\hline
 HIP     &$T_{\rm eff}$   &$\varpi \pm \sigma_\varpi$ &$M_{\rm bol}$&$m_1$   &$m_1^{\tt inf}$&$m_1^{\tt sup}$&$m_2 \pm \sigma_{m2}
$& $q \pm \sigma_q $ \\ 
\hline
 443	 & 4750	 & 24.7	$\pm$0.98 & 0.99    & 	1.7  & 1.3  & 2.1  & 0.76$\pm$0.11  & 0.45$\pm$0.05\\
 2170	 & 4660	 & 3.65	$\pm$0.94 & 0.41    &   2.   & 1.7  & 2.3  & 1.8 $\pm$0.2   & 0.9 $\pm$0.05\\
 8833	 & 4930	 & 18.37$\pm$0.76 & 0.46    &   2.3  & 2.   & 2.6  & 0.54$\pm$0.04  & 0.23$\pm$0.01\\
 8922	 & 4270	 & 9.66	$\pm$0.83 & 0.45    &   1.   & 1.   & 1.2  & 0.7 $\pm$0.05  & 0.62$\pm$0.03\\
 10340	 & 3960	 & 6.0	$\pm$0.74 & -2.33   &   2.   & 1.7  & 2.7  & 0.5 $\pm$0.1   & 0.25$\pm$0.03\\
 10366	 & 4900	 & 16.15$\pm$0.81 & 0.88    &   2.   & 1.6  & 2.4  & 1.12$\pm$0.13  & 0.56$\pm$0.05\\
 10514	 & 4030	 & 3.19	$\pm$0.82 & -2.17   &   2.   & 1.5  & 2.5  & 1.44$\pm$0.2   & 0.72$\pm$0.08\\
 16369	 & 4710	 & 8.955$\pm$0.88 & -1.59   &   4.   & 3.3  & 4.7  & 1.13$\pm$0.13  & 0.28$\pm$0.03\\
 46893	 & 4440	 & 7.72	$\pm$0.90 & 0.07   &   1.5  & 1.2  & 2.   & 0.54$\pm$0.1   & 0.36$\pm$0.04\\
 52085	 & 5000	 & 15.61$\pm$0.77 & 0.47    &   2.5  & 2.1  & 2.9  & 0.52$\pm$0.06  & 0.21$\pm$0.01\\
 54632	 & 5910	 & 22.82$\pm$0.58 & 3.77    &   1.15 & 1.0  & 1.3  & 0.85$\pm$0.04  & 0.74$\pm$0.03\\
 57791	 & 4670	 & 12.67$\pm$1.00 & 0.62    &   1.7  & 1.2  & 2.2  & 0.90$\pm$0.17  & 0.53$\pm$0.07\\
 59459	 & 4740	 & 5.37	$\pm$0.84 & -0.13   &   2.5  & 2.1  & 2.9  & 0.72$\pm$0.07  & 0.29$\pm$0.02\\
 59856	 & 4500	 & 10.45$\pm$0.91 & -0.50   &   2.   & 1.7  & 2.6  & 0.62$\pm$0.11  & 0.31$\pm$0.03\\
 61724	 & 4840	 & 11.69$\pm$1.00 & 0.36    &   2.3  & 2.0  & 2.8  & 0.82$\pm$0.1   & 0.36$\pm$0.03\\
 65417	 & 4660	 & 5.	$\pm$0.85 & -0.60   &   2.7  & 2.2  & 3.2  & 2.5 $\pm$0.25  & 0.92$\pm$0.07\\
 69112	 & 4120	 & 6.26	$\pm$0.47 & -2.07   &   2.5  & 2.   & 3.   & 1.74$\pm$0.2   & 0.7 $\pm$0.06\\
 69879	 & 4730	 & 14.63$\pm$0.68 & 0.11    &   2.2  & 1.9  & 2.5  & 0.98$\pm$0.08  & 0.44$\pm$0.03\\
 83575	 & 4600	 & 9.68	$\pm$0.92 & 0.43    &   1.7  & 1.2  & 2.2  & 0.87$\pm$0.16  & 0.5 $\pm$0.08\\
 87428	 & 4530	 & 9.66	$\pm$1.66 & 0.46    &   1.5  & 1.   & 2.3  & 0.74$\pm$0.21  & 0.5 $\pm$0.08\\
 90659	 & 5150	 & 8.43	$\pm$0.87 & 2.35    &   1.6  & 1.2  & 2.   & 0.35$\pm$0.06  & 0.22$\pm$0.02\\
 91751	 & 4800	 & 13.26$\pm$0.08 & 1.11    &   1.7  & 1.6  & 1.8  & 0.71$\pm$0.025 & 0.42$\pm$0.01\\
 92512	 & 4400	 & 10.28$\pm$0.42 & -1.02   &   2.5  & 2.   & 2.8  & 1.3 $\pm$0.16  & 0.54$\pm$0.05\\
 92818	 & 5160	 & 6.535$\pm$0.63 & -1.67   &   4.   & 3.6  & 4.4  & 1.88$\pm$0.11  & 0.47$\pm$0.02\\
 93244	 & 4740	 & 21.74$\pm$0.70 & 0.09    &   2.1  & 1.9  & 2.5  & 0.47$\pm$0.05  & 0.22$\pm$0.01\\
 95066	 & 4940	 & 8.92	$\pm$0.48 & -0.80   &   3.2  & 3.   & 3.4  & 1.4 $\pm$0.05  & 0.44$\pm$0.01\\
 103519	 & 4920	 & 12.29$\pm$0.63 & 0.52    &   2.3  & 2.   & 2.6  & 0.87$\pm$0.07  & 0.38$\pm$0.02\\
 114421	 & 4800	 & 18.25$\pm$0.78 & 0.69    &   2.   & 1.6  & 2.4  & 0.75$\pm$0.1   & 0.38$\pm$0.03\\
\hline
\end{tabular}
\end{table*}

\begin{figure}[htb]
\resizebox{\hsize}{!}{\includegraphics{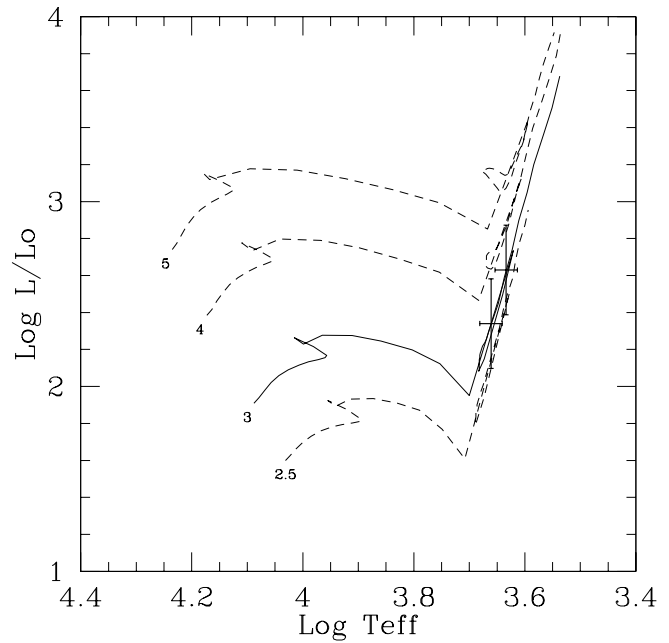}}
\caption[]{\label{fig:sb2hr} Position of the two components of the SB2
HIP 30501 in the H.-R. diagram. The curves are the \citet{Schaller-1992} 
evolutionary tracks, labeled with the initial stellar mass on the main sequence.}
\end{figure}

\begin{figure}[htbp]
\resizebox{\hsize}{!}{\includegraphics{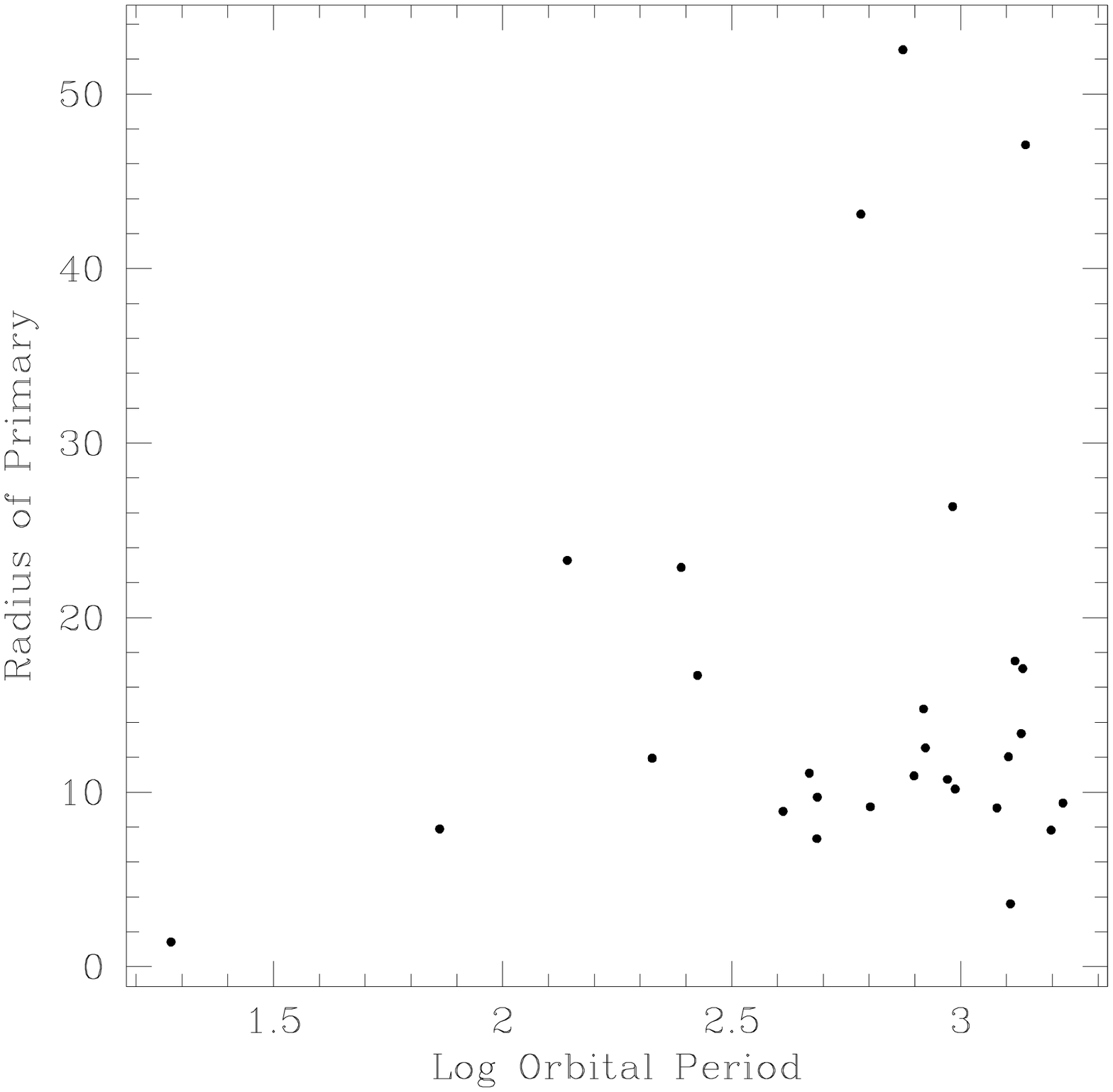}}\\
\resizebox{\hsize}{!}{\includegraphics{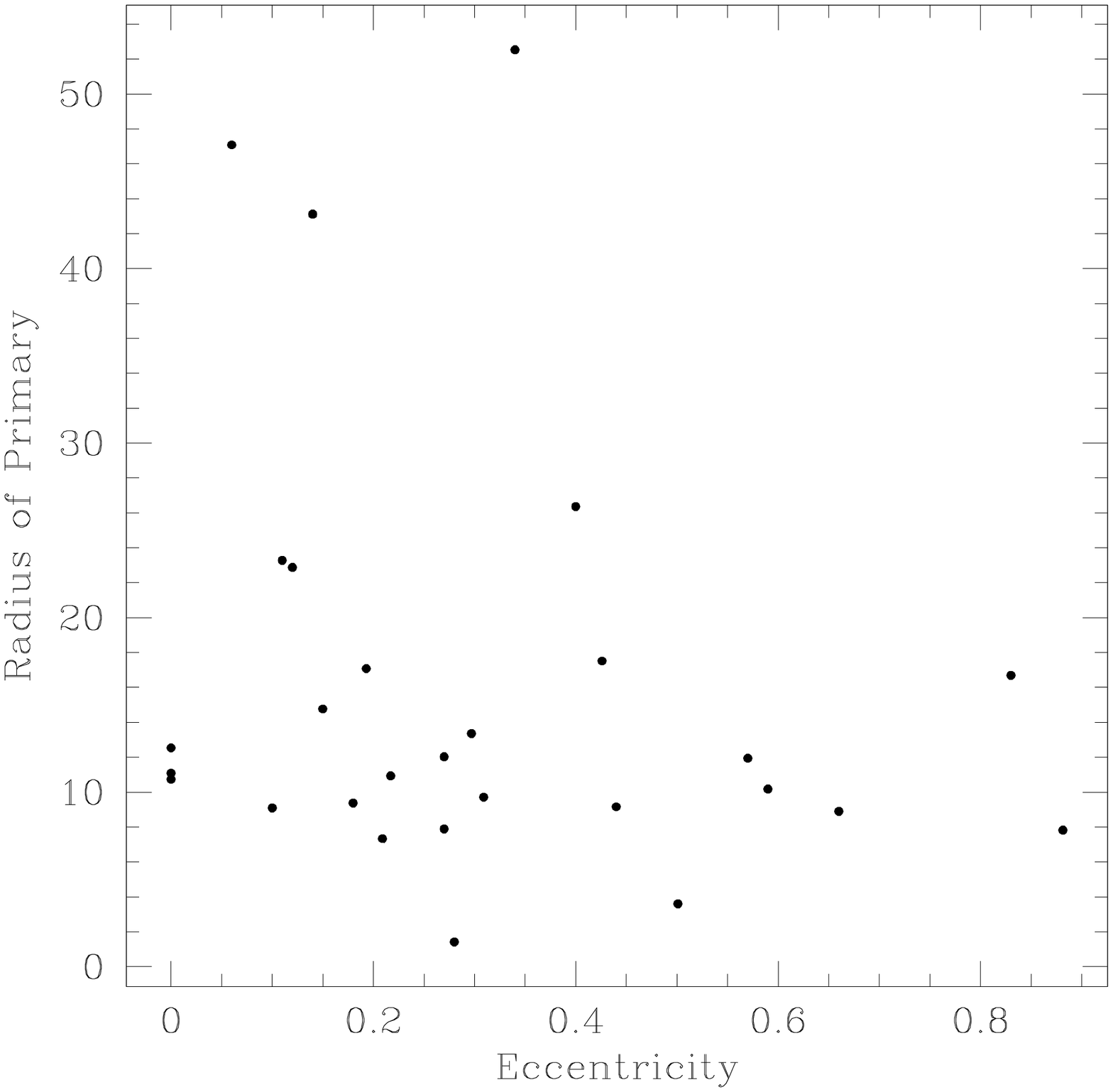}}
\caption[]{\label{fig:radius} 
Radius of the primary of the 28 systems plotted as a function of the logarithm
of the orbital period (above) and of the eccentricity (below). The shortest system has a main sequence primary.
}
\end{figure}

%
\section{Radius and luminosity ratio}\label{Sect:radius}
%
With the value of the luminosity determined and the effective temperature 
known, we can estimate the radius of the primary of the 28 SB1 of our sample.
We plot in Fig.~\ref{fig:radius} the radius as a function of the orbital period 
and of the eccentricity. The first plot clearly shows that the smallest system
contains a main-sequence primary. The smallest period in our sample is thus 73 days, but even for this system, it can be seen from the analysis presented in 
the above sections that the primary is most probably
still in the sub-giant phase. This diagram thus clearly shows the fingerprint
of stellar evolution. It may also show the effect of binary evolution, as the
largest radius are only present in the longer system. 
Three systems have a zero eccentricity and they all have rather large periods.
Unless these systems were formed in a circular orbit, it might well be that,
as explained by BCP93, they either contain  a white dwarf or they are clump
giants which went already through their maximum radius on the first ascent
giant branch. Note that because of the large secondary mass we deduce for
HIP 8922, it is unlikely to contain a white dwarf. 

Once the mass of the secondary has been deduced, and assuming it to be on the 
main-sequence, one can estimate its luminosity, hence the
luminosity ratio between the components. This could then be used in principle
to determine the motion of the photocenter and compare it to the value derived
from the astrometry. This is hardly useful in practice. Indeed, for most systems, the $a_0$ deduced from the astrometric solution has too large an error to 
be used as an additional constraint on the solution. 
Moreover, in most cases, the luminosity ratio is so high that the photocenter is at the location of the primary.
In a few cases, however, this can be used as a check of the correctness of our 
analysis. For example, for HIP 8922, we can deduce from our solution that the
luminosity ratio is 5.5~10$^{-3}$. We can also compute that the semi-major axis is 2 AU,
or 19.6 mas. Given the mass ratio of 0.7, we predict a motion of the
photocenter of 8.1 mas, exactly the value which comes out from our astrometric
solution. 

As an example of the other extreme, let us consider HIP 2170. From the value of
$i, m_1, m_2$ and $P$, we can deduce a separation about 3 AU, or 11 mas. With the
$a_0$ from the astrometry ($6 \pm 2$ mas) and the value of $m_2/(m_1+m_2)=0.47$, 
it appears that the luminosity ratio must be very small. However, from the value of $m_2$, 
one expects a luminosity ratio of about 20\%. Thus, either the secondary 
is itself a close binary or the error on $a_0$ preclude from any conclusion.
There is however enough place in this system to have a hierarchical triple
with the inner binary having an orbital period of 40 days or less.

A few other examples are discussed individually in Sect.~\ref{Sect:individual}.

%
\section{HIP 30501: a SB2 with an astrometric solution}\label{Sect:sb2}
%
HIP 30501 is the only double-lined spectroscopic binary which was kept from our initial full sample.  This is hardly surprising as, for an SB2, the luminosity ratio being close to 1, the photocenter's motion is very much reduced with respect to the one of the binary components and is therefore very difficult to detect.  It is thus quite a surprise to nevertheless still believe that in one case Hipparcos detected the astrometric motion of the photocenter. 

HIP30501 is a 577 days binary with a mass ratio of 0.97 \citep{Griffin-1986}. It contains two giants in an $e=0.24$ eccentric orbit. The solution we derive for this system is different, but within the errors bars, from the one listed in the DMSA/O : $i = 109 \pm 12 \degr $ (instead of $ 121 \pm 21 \degr$); $a_0 = 4.16 \pm 1.28 $ mas (instead of $2.92 \pm 1.13$ mas; $\varpi = 3.36 \pm 0.94$ mas (instead of $3.92 \pm 0.96$ mas.

With our value of $i$ - or the one of Hipparcos for that matter - no eclipse should unfortunately occur in this system.  Using $i$, we derive the masses of the two components: $m_1 = 3.10 \pm 0.65$ M$_\odot$ and $m_2 = 3.02 \pm 0.64$ M$_\odot$. 

\citet{Griffin-1986} estimates the difference in magnitude as $1.24 \pm 0.10 $ mag if the two components have the same spectral type.  It is this difference in magnitude which explains why Hipparcos could detect the motion of the photocenter. Would the two stars have been closer in their luminosity, the detection 
would have proven impossible.  The relatively large uncertainty on $a_0$, $\sin i $ and $\varpi$ makes it unfortunately not possible to use the value of $a_0$ to constraint the luminosity ratio.

A note of caution is here required. 
If the 1.24 mag difference is real then, because the mass of the two stars are very close and because they are on the steep slope of the giant branch in the H-R diagram, they cannot in fact have the same effective temperature!  And therefore one cannot believe the 1.24 mag estimate.  In reality, there is a range of solution possible, which is limited by the fact that the integrated $(B-V)$ index be equal to 1.21 and by the value of the absolute visual magnitude of the whole system. This allows the primary to be of spectral type K2-K3 III, while the secondary would be of type K0-K1 III. The difference in magnitude ranges from 0 (if both star have the same spectral type, a rather unlikely situation has they have different masses but the same age) to about 1.4 (when the primary would become too cool and too bright to fulfill the constraints). These parameters imply an age of roughly $4.~10^8$ years, based on the \citet{Schaller-1992} evolutionary tracks.  We show one of the possible solution in Fig.~\ref{fig:sb2hr}.

It has to be noted that 3 of our 28 single-lined spectroscopic binaries might in fact be double-lined as the estimate of their difference in magnitude is about 1.1-1.3 if one assumes the companion is a main-sequence star of mass $m_2$.  These three stars are HIP 2170, 54632 and 65417.  The fact that there were not detected as SB2 yet might be due to observations done at too low resolution (e.g. HIP 54632), with a radial-velocity spectroscopy not sensible enough to A-F type star (e.g. HIP 2170) or that the companion itself is a close binary (e.g. HIP 65417).
Further studies of these stars are thus called for.

%
\section{Individual cases}\label{Sect:individual}
%
In this section, we now look in turn to each individual cases.
\begin{description}
\item{HIP 443:} 
This is a RS CVn star with an orbital period below 73 days. It is rather surprising that Hipparcos could detect the astrometric signature of this system, but this is certainly due to the relatively large value of the parallax. Although the DMSA/O solution quoted a value of $36 \pm 37 \degr$ (!) for the inclination, our reanalysis shows a more secure value of $44 \pm 12 \degr$. The secondary is constrained to be a low mass star, in the range 0.5 to 1 M$_\odot$.  Although there were several attempts, this system has never been resolved by (speckle) interferometry \citep{Hartkopf-2001:b}.

\item{HIP 2170:}
One of the systems with a rather large mass ratio, between 0.8 and 1. The parallax we deduce is sensibly different from the one in the Hipparcos catalogue (3.65 vs. 1.61). 

\item{HIP 8833:}
We obtain a slightly larger value of the parallax than the Hipparcos DMSA/O solution and a more accurate orbital solution. \citet{Zhao-2001:a} quote a value of 2.5 M$_\odot$ for the mass of the giant, in good agreement with our $2.3 \pm 0.3$ M$_\odot$ value. The mass ratio is well constrained to lie around 0.23, leading to a companion mass of 0.55 M$_\odot$. No enhancement of s-process elements was detected by \citet{McWilliam-1990:a}.  Although there were several attempts, this system has never been resolved by (speckle) interferometry \citep{Hartkopf-2001:b}.

\item{HIP 8922:} 
Although this system has an orbital period of 838 days, Hipparcos did not find an orbital solution. Our reanalysis however provides a very precise value for the inclination, $23 \pm 2 \degr$.  With such a large period, a null eccentricity and a very small mass function, this star belongs to the class of stars defined by BCP93 to possibly harbor a white dwarf. The value for the secondary mass, $0.7 \pm 0.05$ M$_\odot$ is certainly not an argument against. Unfortunately, no abundance analysis has been published for this star so that we cannot check if it belongs to the class of PRGs.  Although there were several attempts, this system has never been resolved by (speckle) interferometry \citep{Hartkopf-2001:b}.

\item{HIP 10340:}
Another long period, low mass function system. The range in mass for the secondary does also allow a degenerate companion. For this star, however, stellar abundance show it not to be enriched in s-process elements \citep{Jorissen-1992:a, McWilliam-1990:a} even though it is sometimes referred as Ba0.5. It might thus represent an example of the fact that binarity is not a sufficient condition for the barium star phenomenon.  
It has however been claimed that 
this system was resolved by speckle interferometry \citep{McAlister-1987} even if only two measurements are available \citep{Hartkopf-2001:b}.
With the parameters we derive, such a detection is questionable.

\item{HIP 10366:}
We derive a very accurate solution for this very eccentric orbit ($e=0.8815$) and deduce a 2 M$_\odot$ mass for the giant and a 1.1 M$_\odot$ for its companion. This star is among the normal stars in the sample of \citet{McWilliam-1990:a}.  Our solution leads to a semi-major axis of 3.87 AU or 62.6 mas, and taking the face values of the components masses, a value of $a_1 = 22.5$ mas in perfect agreement with the value obtained from spectroscopy.
This is close to the $a_0$ value derived from astrometry, even though it is a little too large, especially if one has to take into account the fact that it is photocentric motion and not the motion from the giant that we see.  \citet{Balega-1984:a} once resolved this system by speckle interferometry.


\item{HIP 16369:}
One of the most massive giant in our sample, with a primary mass of $4 \pm 0.7 $ M$_\odot$ and a companion with a mass close to 1 M$_\odot$. Although the DMSA/O solution quotes a value of the inclination of $37 \pm 29 \degr$, our reanalysis provides a very different but well determined value of $66 \pm 8 \degr$.  Although there were several attempts, this system has never been resolved by (speckle) interferometry \citep{Hartkopf-2001:b}.

\item{HIP 46893:}
This star was in the DMSA/X annex of the Hipparcos catalogue which could not find an orbital solution. Our analysis provides it with a very similar but more accurate parallax.  Although there were several attempts, this system has never been resolved by (speckle) interferometry \citep{Hartkopf-2001:b}.

\item{HIP 52085:}
The parallax we obtain is slightly larger than the one in the Hipparcos catalogue. Our orbital solution also is different from the one in the DMSA/O with an inclination of $128 \pm 4 \degr$ instead of $147 \pm 12 \degr$.  \citet{McWilliam-1990:a} does not detect any s-process enhancement in this star.  Although there were several attempts, this system has never been resolved by (speckle) interferometry \citep{Hartkopf-2001:b}.

\item{HIP 54632:}
The shortest orbital period in our sample, 19 days, and identified as single in the Hipparcos catalogue.  This is clearly a main-sequence star of spectral type G0 V and a mass of $1.15 \pm 0.1 $ M$_\odot$. The mass ratio is in the range 0.5 to 0.9, even though the largest value must not be the correct one as the system would appear as an SB2 in this case. 

\item{HIP 57791:}
One of the few example where our solution is similar to the one given in the DMSA/O annex.  With an inclination of $86 \pm 6 \degr$, this system could be eclipsing. The rather secure mass ratio we derive is $0.53 \pm 0.07$.  This system is resolved by speckle interferometry \citep{McAlister-1983:b} but only three measurements have ever been obtained \citep{Hartkopf-2001:b}.

\item{HIP 59459:}
No luminosity class exist for this object but our analysis shows it to be a giant albeit slightly hotter than the K2 spectral type mentioned in the Simbad database.  

\item{HIP 59856:}
\citet{McWilliam-1990:a} does not detect s-process enhancement in this star for which we obtain a companion mass around 0.6 M$_\odot$ and a mass ratio close to 0.31. Our solution is more accurate and in better agreement with the spectroscopic one than the Hipparcos DMSA/O one.  Although there were several attempts, this system has never been resolved by (speckle) interferometry \citep{Hartkopf-2001:b}.

\item{HIP 61724:}
Another case of a possible eclipsing system and a solution in agreement with the one quoted in the DMSA/O.  Although there were several attempts, this system has never been resolved by (speckle) interferometry \citep{Hartkopf-2001:b}.

\item{HIP 65417:}
This is the binary with the larger mass ratio we found, $q = 0.92 \pm 0.07$. Assuming a main sequence companion, the difference in visual magnitude could therefore only be slightly larger than 1 magnitude (but see Sect.~4.). 
The star was observed by \citet{Zacs-1997:a} has having no s-process enrichment, which is no surprise if as we estimate the companion must have a mass above 2 solar masses.  Although there were several attempts, this system has never been resolved by (speckle) interferometry \citep{Hartkopf-2001:b}.

\item{HIP 69112:}
We find a giant mass of $2.5 \pm 0.5$ M$_\odot$ and a mass ratio around 0.7. The companion, with a mass close to 1.7 M$_\odot$, if on the main-sequence, must be of spectral class A7 V. Unpublished IUE observations of this star \citep{Boffin-1993:c} show indeed the presence of a late A-early F companion.  \citet{Richichi-2002:a} quote an angular diameter for the giant of 2.76 mas. With our adopted  solution, we found a value of 80 solar radii, or 2.41 mas, close to the above-mentioned value.  \citet{Bonneau-1986:a} resolved this system but other attempts remained unsuccessful \citep{Hartkopf-2001:b}.

\item{HIP 69879:}
This is certainly, with an inclination of $90 \pm 9 \degr$, one of the best candidate of an eclipsing system. The solution we found is also well constrained with a primary mass of $2.2 \pm 0.3$ M$_\odot$ and a mass ratio of $0.44 \pm 0.03$. Taking the parameters of the K0 III giant, i.e. a radius of 12.3 R$_\odot$ and assuming a one solar mass companion, one obtain an eclipse depth of 9 mmag. In the Hipparcos catalogue, HIP 69879 is classified as a new variable with an amplitude of 17 mmag. However, we have checked that no periodicity appear in the Hipparcos data and the variability detected is therefore not consistent with the assumption of an eclipse.  Although there were several attempts, this system has never been resolved by (speckle) interferometry \citep{Hartkopf-2001:b}.

\item{HIP 83575:}
Our solution for this star is similar, but more accurate, than the one quoted in the DMSA/O. The mass ratio is very close to 0.5.  Although there were several attempts, this system has never been resolved by (speckle) interferometry \citep{Hartkopf-2001:b}.

\item{HIP 87428:}
Noted as single in the DMSA/O, the solution we obtain is not very useful as the uncertainty on the mass function is of 100 \%. 

\item{HIP 90659:}
A G8 III-IV which is indeed verified by our solution, not very different from the DMSA/O one. \citet{Zacs-1997:a} do not detect any s-process enhancement and our analysis show that the companion is most probably a low mass star, in the range 0.3-0.4 M$_\odot$. This is too small a value for a white dwarf. 

\item{HIP 91751:}
Here again, our solution is close to the DMSA/O but slightly more accurate. The mass ratio of this system is well constrained in the $0.42 \pm 0.03$ range.  Although there were several attempts, this system has never been resolved by (speckle) interferometry \citep{Hartkopf-2001:b}.

\item{HIP 92512:}
Another potential eclipsing system, also quoted as variable in the Hipparcos catalogue with an amplitude of 0.04 mag. With its rather short period of 138.7 days, it is however of the RS CVn type which could be the cause of the variability.  Our solution implies a 1.3 M$_\odot$ companion and a 2.5 M$_\odot$ primary. An eclipse should have an amplitude of about 3-4 mmag only.  Although there were several attempts, this system has never been resolved by (speckle) interferometry \citep{Hartkopf-2001:b}.

\item{HIP 92818:}
The second 4 solar mass giant in our sample with a mass ratio slightly below 0.5. With our solution, we derive a radius for the G4 III giant of 24 R$_\odot$ or 0.73 mas. This is in very good agreement with \citet{Blackwell-1991:a} who found an angular diameter of 1.4 mas.  \citet{Hummel-1995} found a A6V companion with a difference of 2 magnitudes between the components.  This is also in very good agreement with the 1.9 M$_\odot$ we derive for the companion, even though we predict a 3.4 mag difference in V.

\item{HIP 93244:}
Another possible eclipsing system, with an inclination of $87 \pm 7 \degr$. Our solution, close to the DMSA/O one, gives a radius of 12 R$_\odot$ (1.2 mas) for the 2 M$_\odot$ giant, in perfect agreement with the value obtained by \citet{Nordgren-1999:a}.  The companion should have a mass between 0.4 and 0.5 M$_\odot$ and an eclipse would therefore only show a dip of 2-3 mmag. Although this star has been sometimes referred as a Ba0.2 star, \citet{McWilliam-1990:a} found a normal s-process abundance and we guess that the companion is a normal low-mass star.  With such a small companion, the difference in magnitude between the two components should be more than 7 magnitudes, and the photocentric motion can therefore very well be approximated by the motion of the giant star. In this case, taking the face value of the masses we derived, we obtain a semi-major axis of 3.14 AU or 68 mas. The value of $a_1$ is therefore about 12.5 mas, in very good agreement with the value deduced from the $a_1 \sin i$ value deduced from spectroscopy and the $a_0$ one deduced from the astrometric solution. We are thus confident about the accuracy of our model.

\item{HIP 95066:} Although there were several attempts, this system has been resolved only twice by (speckle) interferometry \citep{Hartkopf-2001:b}.

\item{HIP 103519:}
Our solution is very similar to the DMSA/O one. \citet{McWilliam-1990:a} did not detect any enhancement of s-process elements in this giant. We derive a mass of the companion close to 0.9 M$_\odot$ for a mass ratio around 0.38.  Although there were several attempts, this system has never been resolved by (speckle) interferometry \citep{Hartkopf-2001:b}.

\item{HIP 114421:}
We obtain a slightly larger value of the parallax than the Hipparcos DMSA/O solution. The mass ratio is here also around 0.38.  Although there were several attempts, this system has never been resolved by (speckle) interferometry \citep{Hartkopf-2001:b}.
\end{description}

%
\section{Conclusion}\label{Sect:conclusion}
%
We have updated the catalogue of spectroscopic binaries containing a red giant of \citet{Boffin-1993:a} and cross-identified it with the Hipparcos catalogue. 
A sample of 215 systems was obtained.  For these, we have reanalyzed their Hipparcos Intermediate Astrometric Data and applied new statistical tests which combine the requirement for the consistency between the Thiele-Innes and the Campbell solution as proposed by \citet{Pourbaix-2001:b} with the requirement that the most significant peak in the Hipparcos periodogram corresponds to the orbital period. 
By doing so, we select 29 systems among which one double-lined spectroscopic binary, for which we are confident the newly derived astrometric solution is consistent with the spectroscopic orbit.  
Among these, 6 are new orbital solutions not present in the DMSA/O catalogue.
On the other hand, our procedure has rejected 25 DMSA/O entries. For some of these, new radial velocities are indeed in contradiction with the DMSA/O orbits, confirming our rejection.

Our sample of 29 systems allows to derive the distributions of the masses of the components as well as the mass ratio distribution. 
We find that the mass of the primary is peaked around 2 M$_\odot$, that the secondary mass distribution is consistent with a Salpeter-like IMF and that the mass ratio distribution rises for smaller value of the mass ratio.

The only double-lined binary in our sample is HIP 30501 for which we find that the primary must be of spectral type 
K2-K3 III, while the secondary would be of type K0-K1 III. Three other systems should be more closely studied (HIP 2170, 
54632 and 65417) as our solution make them potential double-lined systems.

Finally, we find that five systems might have inclination close to 90 degrees (HIP 57791, 61724, 69879, 92512 and 93244)
and could therefore possibly show eclipses. The depth of which will typically be of the order of 0.01 mag or below.

\acknowledgement{This research was supported in part by ESA/PRODEX 14847/00/NL/SFe(IC) and C15152/01/NL/SFe(IC).
It is a pleasure to thank the referee, F. Arenou, for useful remarks on the paper.}

\bibliographystyle{aa}
\bibliography{articles,books}
\end{document}